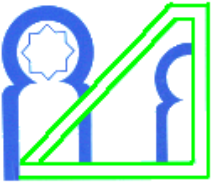

UNIVERSITÉ MOHAMMED V – AGDAL
ECOLE MOHAMMADIA D'INGENIEURS
RABAT

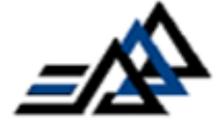

Centre d'Etudes Doctorales
Sciences et Techniques pour l'Ingénieur

# Thèse de Doctorat

Présentée et soutenue publiquement par

## Youssef GAHI

Pour obtenir le grade de : **Docteur en Sciences et Techniques pour l'Ingénieur**

Structure de recherche : **Laboratoire d'Electronique et Communications**

Spécialité: **Informatique**

---

# On the use of homomorphic encryption to secure remote computing, services, and routing protocols

---

Soutenue le, 01 Octobre 2013, devant le jury composé de:

| | | |
|---|---|---|
| **Pr. Belhaj ELGRAINI** | **Président** | **EMI** |
| **Pr. Dalila CHIADMI** | **Rapporteur** | **EMI** |
| **Pr. Fouzia OMARY** | **Rapporteur** | **FSR** |
| **Pr. Mohamed ERRADI** | **Rapporteur** | **ENSIAS** |
| **Pr. Zohra BAKKOURY** | **Examinatrice** | **EMI** |
| **Pr. Zouhair GUENNOUN** | **Directeur de thèse** | **EMI** |
| **Pr. Mouhcine GUENNOUN** | **Co-directeur de thèse** | **UOIT** |

# Abstract

The trend towards delegating data processing and management to a remote party raises major concerns related to privacy violations for both end-users and service providers. These concerns have attracted the attention of the research community, and several techniques have been proposed to protect against malicious parties by providing secure communication protocols. Most of the proposed techniques, however, require the involvement of a third party, and this by itself can be viewed as another security concern. These security breaches can be avoided by following an innovative approach that depends on data sorted, managed, and stored in encrypted form at the remote servers. To realize such an approach, the encryption cryptosystem must support algebraic operations over encrypted data. This cryptosystem can be effective in protecting data and supporting the construction of programs that can process encrypted input and produce encrypted output. In fact, the latter programs do not decrypt the input, and therefore, they can be run by an un-trusted party without revealing their data and internal states. Furthermore, such programs prove to be practical in situations where we need to outsource private computations, especially in the context of cloud computing. Homomorphic cryptosystems are perfectly aligned with these objectives as they are a strong foundation for schemes that allow a blind processing of encrypted data without the need to decrypt them. In this dissertation we rely on homomorphic encryption schemes to secure cloud computing, services and routing protocols. We design several circuits that allow for the blind processing and management of data such that malicious parties are denied access to sensitive information. We select five different areas to apply our models to. These models are then easily customized for many other areas. We also provide prototypes that we use to study the performance and robustness of our models. Finally, we set our objectives for future research directions.

# Acknowledgements

This thesis has been prepared within collaboration between, École Mohammadia d'Ingénieurs (Morocco), and University of Ontario Institute of Technology (Canada).

First and Foremost, I would like to thank my supervisors Pr. Zouhair Guennoun, Pr. Mouhcine Guennoun, and Pr. Khalil El-Khatib, for their confidence, support, time, guidance, detailed feedback, and encouragement through the different steps of this thesis. Also, I would like to provide special thanks to Pr. Mouhcine Guennoun for his special efforts and all advices that he provided during this project.

A very special thanks goes out to my parents and my family, for their unconditional help, support, time, and above all, understanding the difficult and busy times that I passed through.

I would also like to thank my friends: Bahia Rached, Yousra Harkat, Zakaria Mohammadi, Anni Driss, and Yacine Ichibane for their continuous moral support and for their ongoing help throughout the course of my studies.

I must also acknowledge members, professors, and students of Laboratory of Electronics and Communications, for their encouragement, help, and professionalism.

I would also like to thank the members of the jury for accepting to judge this thesis.

Finally, I would like to thank all those who believed I could achieve this special dream...

# Table of Contents

# List of Tables

# List of Figures

# List of Acronyms

| | |
|---|---|
| ASCII | American Standard Code for Information Interchange |
| FHES | Fully Homomorphic Encryption Scheme |
| SHES | Somewhat Homomorphic Encryption Scheme |
| SMM | Secure Meta Mediator |
| RAM | Random Access Memory |
| VOD | Video On-Demand |
| AVG | Average |
| ACR | Availability-Centric Routing |
| DCT | Discrete cosine transform |
| DCT | Discrete Cosine Transform |
| SQL | Secure Query Language |
| LBS | Location-Based Service |
| GPS | Global Positioning System |
| HES | Homomorphic Encryption Scheme |
| FHE | Fully Homomorphic Encryption |
| TLS | Transport Layer Security |
| LTE | Long-term-evolution Advanced |
| API | Application Programming Interface |
| SSS | Secret Sharing Scheme |
| RR | Route Request |
| RP | Route Reply |
| ID | Identifier |
| IT | Information Technology |

# List of Symbols

$\varepsilon_{pk}(m)$    The encryption result of a bit m under the public key pk

Wi-Fi         device to exchange data wirelessly

ACK           Acknowledgement

XOR           A logical Gate to perform addition

AND           A logical Gate to perform multiplication

add           Addition

mult          Multiplication

$c^+$         The re-encrypted form of bit value m

pk            The public key used to encrypt a plain bit value.

sk            The secret key used to decrypt a ciphertext

C             A circuit to operate successfully on encrypted values

T             The database table

P             Product of two big primes

*             Star Gate

$\lambda$     The level of security of the homomorphic scheme

c             The encrypted form of a bit value m             the plain bit value.

q             The parameter of security in FHE scheme

r             The noise term in the encrypted value

# Chapter 1

# Introduction

## 1.1  Background

The popularity of public networks, especially the Internet, and the unprecedented growth in their users has raised major security concerns, especially in terms of users' privacy. The risks of violating users' privacy exacerbate when we consider the growing trend towards using smart devices, the emergence of cloud services, and the interest in mobility solutions. These trends require full collaboration between clients, service providers and web hosts to form a solid foundation for a secure environment. This environment should implement sufficient security measures that can prevent intruders from performing a diverse range of malicious actions. Among the known approaches to support security are the usage of encryption/decryption techniques, key sharing, third party delegation, and blind processing. These approaches aim at hardening the communication system by combining cryptographic theory and secure computations. Having a secure communication system paves the way for a huge body of applications that require high levels of privacy preservation, like bank transactions and medical applications. In other words, taking security aspects into consideration when designing any communication system, reflects into high trust in these systems, and encourages people to utilize them.

## 1.2 Motivation

Since the publication of the RSA encryption scheme in 1978, computations on encrypted values have attracted a valuable attention. This type of encryption is an effective approach for data protection since processors do not need to decrypt inputs before performing the requested operations. However, most of the proposed protection schemes that aim at extending the homomorphic encryption basic model were suffering from a lack of flexibility, and have supported only a limited number of arithmetic operations. Recently, a Fully Homomorphic Encryption Scheme (FHES) has been proposed by Gentry to support an exhaustive number of operations with high flexibility [GEN09a]. This novel encryption/decryption method not only help preventing malicious parties from accessing private information, but also makes it possible to develop a rich collection of practical secure applications. We adopt FHE schemes in this thesis to develop strong foundations for different kind of applications. Although innovative in their concept, these schemes are still time consuming and show a modest performance. We are motivated by the fact that these schemes are under the scope of extensive research efforts to improve their performance. We expect that these shortcomings can be bypassed given the advancements achieved in hardware capabilities.

## 1.3 Objectives

The objective of this thesis is to develop foundations and design generic circuits, easily customizable, that can effectively preserve the privacy and confidentiality in various applications. These applications span over five areas of research:

1. Database and applications.
2. Processors and remote program execution.

3. Location Based Services and mobility solutions.

4. Ad-hoc wireless networks and their dynamic collaborations.

5. Video streaming through public networks and secure purchasing protocols.

## 1.4 Contributions

The main research contributions of this thesis are as follows:

1. We have re-designed basic database queries using the Homomorphic Encryption Scheme (HES). This is accomplished through defining circuits as well as employing a novel obfuscation technique to service requests appropriately. This has enabled us to establish a secure collaborative mode with which queries are blindly processed without giving unauthorized parties any access to the data of interest.

2. We have presented an encrypted processor that allows executing programs without revealing their details and internals. This processor is based on a novel logical gate called the Star. The Star protects the program content by hiding the gates that are being executed at a specific time instant. This means that our novel processor can be forced to execute binaries without the need to disclose any information.

3. We have built a secure architecture for Location Based Services to demonstrate the effectiveness of our aforementioned novel work. We have used FHES to develop circuits that can process client queries anonymously, and find the requested point of interests without knowing the exact locations of those users. Moreover, our proposed solution can combine plain and encrypted records such that servers do not need to encrypt their databases to support such a topology. On the other hand, the results returned to the requesting client will be implicitly encrypted by performing operations between plain and cipher text. That is, only the requester can decrypt the returned locations. Our architecture can be easily extended to serve other applications (especially in the context of mobility solutions).

4. We have developed a secure mechanism for information exchange in ad-hoc wireless networks. These networks suffer from major security vulnerabilities due to their dynamic nature. We have proposed circuits that allow nodes to share their experience, and evaluate their in-neighbors without disclosing any confidential information publicly. This way a secure routing process can be guaranteed and trusted routes can be established without involving any malicious or suspicious nodes over those routes.

5. We have developed a fully secure mechanism that establishes a secure protocol between video requesters and service providers. This protocol allows each party to collaborate with the other without violating their privacy. This protocol answers the increased demands for video streaming and the lack of security in the available systems to support this popular application.

## 1.5 Thesis Outline

The remainder of this thesis is organized as follows. In Chapter 2, we show main reasons of loss of privacy in Internet as well as how to provide a secure use. Then, in Chapter 3, we present the scheme adopted in our work, and survey related works that aim at securing sensitive applications. In Chapter 4, we describe our proposal to re-design and protect SQL queries such that a secure database system is achieved. Chapter 5 details our novel technique for blind program execution. In Chapter 6, we propose an efficient secure solution to support Location Based Services. Next, within Chapters 7 and 8 we provide a description for circuits designed to achieve secure routing for ad-hoc wireless networks and streaming protocols, respectively. Finally, Chapter 9 concludes our thesis report and provides future research directions.

## 1.6  List of Publications

### 1.6.1 Articles in Referred Journals

1.  Y. Gahi, M. Guennoun, Z. Guennoun, and K. El-khatib. Privacy Preserving Scheme for Location-Based Services. In *Journal of Information Security*, Vol. 3, No. 2, 2012.

2.  Y. Gahi, M. Guennoun, Z. Guennoun, and K. El-khatib, "On the use of homomorphic encryption to secure applications, services, and routing protocols," In *The European Journal Of Scientific Research*, Vol. 88, No. 3, 2012.

3.  Y. Gahi, M. Guennoun, Z. Guennoun, and K. El-khatib, " Securing Internet Applications using Homomorphic Encryption Schemes," In *The Journal of Theoretical and Applied Information Technology*, Vol. 47, No. 1, 2013.

### 1.6.2 Articles in Referred Conference Proceedings

4.  Y. Gahi, M. Guennoun, and K. El-khatib, "A Secure Database System using Homomorphic Encryption Schemes," In *the 3rd Int. Conf. on Advances in Databases, Knowledge, and Data Applications*, pp. 54-58, St. Maarten, The Netherlands Antilles, 2011.

5.  Y. Gahi, M. Guennoun, Z. Guennoun, and K. El-khatib, "Encrypted Processes for Oblivious Data Retrieval", In *The 6th IEEE Int. Conf. for Internet Technology and Secured Transactions*, pp. 514-518, Abu Dhabi, 2011.

6.  Y. Gahi, M. Guennoun, Z. Guennoun, and K. El-khatib, "A Fully Private Video on-Demand Service," In *The 25th IEEE annual Canadian Conference on Electrical and Computer Engineering*, pp. 1-5, Montreal, Canada, 2012.

# Chapter 2

# The Loss of Privacy

## 2.1  Introduction

In its beginning, the Internet was mainly introduced to interconnect systems and laboratories that are committed to government research [LEI09].

As years passed, the main goal of using the Internet has expanded to allow billions of users to connect with one another regardless of their affiliations, see Figure2-1. Hence, nowadays, the Internet is referred to as the worldwide interconnection of people over individual networks that can be led by private parties, industries, or governments.

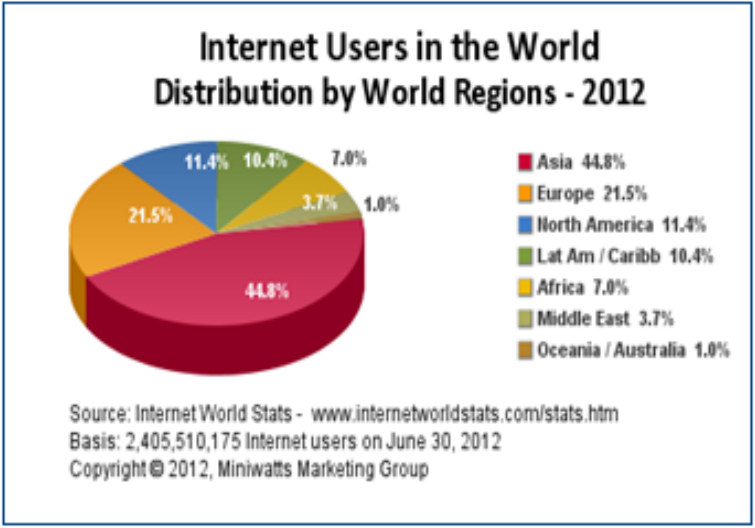

Figure 2-1: Internet usage by regions [1].

This tremendous growth played an important role in making computer-based information systems, or even more specifically online-based systems, ubiquitous [VEE04]. This means that these elements are highly used in both individuals' and businesses' daily transactions. It is actually noteworthy that this technological revolution brought many advantages that can be as simple as easing the process of communication between people, or as complex as handling huge e-commerce companies' day-to-day operations. However, its drawbacks can be more harmful than individuals can ever expect. One of the most dangerous threats is the loss of online privacy and confidentiality [GHO13].

In the digital age, Internet privacy refers to many concepts depending on people's perceptions. Some of us are more concerned about controlling their social networks, others are more anxious about the extent to which companies track their behavior whenever they browse different websites, while some people may even worry about governments monitoring their online activities. In other words, Internet privacy entails two major aspects:

- The personal information that identifies an individual such as their home address or phone number, and

- The non-identifying information like the specific behavior of people while surfing the Internet.

## 2.2  Importance of Internet Privacy

Online privacy is actually a sensitive matter because of the increasing number of Internet users as well as the threats that menace them [2]. It is important to be aware of the existence of a large number of hackers and stalkers that aim at violating the users' privacy and benefit from an unauthorized access to their undisclosed personal information. This unwanted behavior can, not only damage one's privacy, but also be a major source of financial and social worries. In order to be convinced why it is extremely important to safeguard Internet privacy, one should be aware of the various risks and dangers that are raised whenever and whatever an unsecure Internet access occurs.

---

[2]    Canadian Internet Policy and Public Interest Clinic report, 2008 [URL03]

It is common to notice that while advertising, websites target Internet users based on their interests. In order to do so, these websites should have access to their data and, sometimes, private information. Some are satisfied because they only see the advertising banners that appeal to them, however, this matter is not fully as innocent as it seems. The real issue is raised when these parties share the information they have with other entities without the users' knowledge or agreement. This leads us to examine more the available Internet privacy issues that threaten people while using the Internet. Some of the privacy threats consist of the unauthorized network access, hacking, phishing, email fraud and spamming. Among these elements, hacking is the most used technique to acquire people's information and harm them [HAT12]. Once a party hacks users' data, they have complete control over how this information can be utilized and can misuse it for illegal or irrational purposes, see Figure2-2 and Figure2-3 for some statistics about web application security and hack techniques as well as the cost of cybercrime attacks for different countries during 2012.

Many examples can illustrate how people are exposed to Internet privacy loss, such as the case when the Internet Service Provider (ISP) can have full access to the users' information simply by tracking their online activities and behavior.

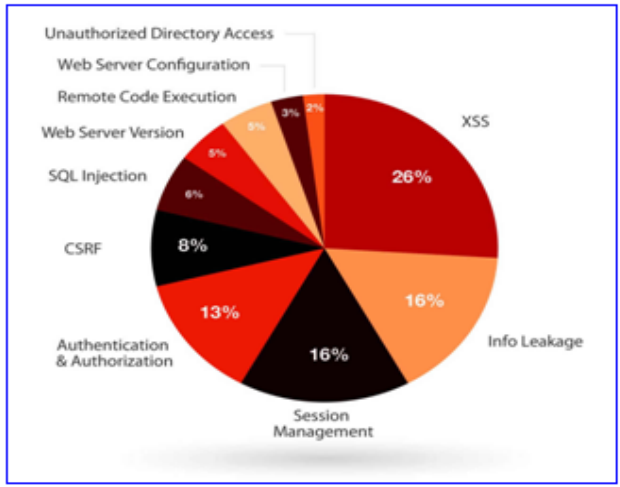

Figure 2-2: 2012 Web Application Security Vulnerability Population [3].

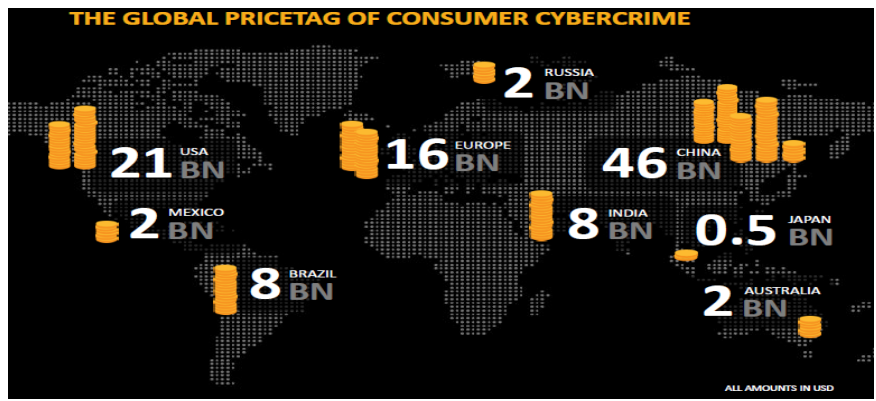

Figure 2-3: The cost of cybercrime for different countries [4].

## 2.3  Causes of Privacy Loss

The growth of Internet along with the improvement of its different features and advantages has heightened many issues regarding online privacy of the users. Particularly, online users' fear

---

includes the propagation of databases, the likelihood of online privacy loss and violation, as well as the misuse of the personal information that has been collected. The general causes that have contributed to the increasing of these fears involve the lack of official and legal frameworks that can play an important role into counteracting the offensive behavior of technology. Moreover, the fact that the increasing media publicity promotes the Internet usage as a flawless tool that can easily be used by people. It is rarely mentioned that different threats are awaiting the surfers of the Internet. In addition to these causes, the most dangerous one remains the fact that not all people are aware of privacy issues. Usually, people do not have the required knowledge that can protect them from falling into the trap of losing all their personal information, or worse, suffering from social and financial implications because of the different existing Internet frauds.

Currently, the approximate overall cost for the different cyber crimes is measured in billions, see Figure 2-3. Even if many institutions have invested in both the development and implementation of different measures that can help improve Internet security, computer and Internet misuse is unfortunately expected to keep being a problem as years pass.

The information system professionals are more anxious regarding the propagation of the unethical and unwanted behavior that may occur while making use of computers that are connected to the Internet. This fear is due to the fact that it can negatively affect both the society and the information system profession.

## 2.4  How to Deal with Online Privacy Loss

Once Internet users build a strong understanding about the dangerous online privacy issues, and are aware of the many threats that menace them, such as the fact that some websites generate their official revenues only by selling private information to other entities, they should be willing to wholeheartedly welcome protection from such threats and risks. No user would want to see their private data spread over the web and have no control over who can view the shared information. That's why online users must be vigilant whenever they are ready to surf the web.

Many perspectives can be considered while viewing Internet privacy protection measures. These perceptions range from the individual's adoption of simple behavior that allow them to control the private data that they share on the Internet and the different policies and regulations that exist, to the complex technological measures that can be acquired [QUI07, RAJ12, LU07]. Combining all these online privacy protection perceptions allows all users to be sure that their private pieces of information are successfully protected to a large degree and less likely to be compromised.

If we examine the behavior of an Internet user regarding what information he/she publishes and shares with the public over the web, we notice that the individual's control plays an extremely important role in protecting his/her privacy. A good behavior, combined with some extra cautions, contributes in effectively safeguarding any private information from malicious entities. Before they even consider sharing their private information on a website, users must be fully aware of its credibility. Within the same context, avoiding opening the links that are received from anonymous sources can spare many privacy loss or troubles because they may contain spyware that secretly collects sensitive data and automatically sends it to external, harmful people.

In order to reduce privacy loss or fraud, many policies and regulations are set to control and supervise the different usages of personal information that is submitted to a given organization, such as the companies that embed the e-commerce as part of their services. In this latter example, the users should be fully aware that some employers have access to their sensitive information while trying to ensure the service that they first requested. However, specific restrictions are set to specify to which degree the user's data can be used. In general, similar websites and organizations clearly state their policies and share them with users. Legal actions can be taken against any organization that does not respect the agreement or the policies about the usage of the user's private information.

Finally, Internet users must know that different technological tools were developed in order to contribute in the process of securing the user's online private data and prevent privacy loss [MAR07, XU12, and GUO02]. There is a countless number of online protection software that put together all the required packages (anti-spyware, firewalls, data encryption functionalities and setting…etc) that help the users preserve their anonymity. Making use of all of these elements may allows the users to protect their sensitive information and prevent any privacy loss issues.

## 2.5  Conclusion

Nowadays, the Internet is considered as the largest information container and the most used means of communication. Promoting the social networking features, emailing and blogging have all effectively and efficiently contributed to easing, moving forward and improving the human being's life. This wonderful resource made the whole world looks like a smaller place where billions of people live. However, with the tremendous growth of the Internet, many issues were raised to harm and threaten the most sensitive information about the users: their privacy!

It is extremely important to be aware of the different dangers that can face the Internet users whenever they try to connect to the external world through a public network. One must be also aware of the different parties that present a potential fraud and should know how to deal with them. These malicious entities can be hackers, stalkers or third party users. They all aim at extracting sensitive and private data probably in order to make malicious use of it.

Finally, all Internet users must know the exact approaches that can help them secure their privacy and make use of them. As explained above, they can either make use of the policies and regulations, install the available software that secure their sensitive data, or simply control the information that they share with the public.

# Chapter 3

# Survey of Related Work

This chapter provides a survey of prior work that is related to the problem we tackle in this thesis. First, we exhibit practical and popular applications/systems and highlight the lack of privacy and confidentiality support in them. This lack of security support contributes to the decrease in the number of users of these applications. Then we study different models and techniques that have been proposed to secure these applications/systems. We shed the light on the strengths and limitations of these proposals and discuss how to enhance them such that they keep pace with the rapid development in technology. Finally, we elaborate on the fully homomorphic encryption concept and demonstrate how it can be utilized to support security effectively. We also discuss our main contribution that aim at involving and customizing this novel encryption technique to realize fully private prototypes that are able to secure real-life applications.

## 3.1 Introduction

In the recent years, Internet services and related applications have experienced a notable development and a massive growth in the number of users, see Figure2-1 for the distribution of users by world regions. These trends have benefited from the enhancements in hardware capabilities and the popularity of smart phones, Wi-Fi networks, GPS, as well as mobility solutions.

Although these systems provide a portfolio of facilities, they are usually accompanied with a set of security risks like, piracy, tracking, and monitoring, and these risks often indicate a violation to the privacy. Therefore, security aspects should be obeyed in order to protect end-users.

Researchers have followed different approaches to secure sensitive applications. Supporting security requires paying careful attention to the following aspects:

1. Confidentiality and Integrity: Data confidentiality is an important aspect of security. It refers to that certain parties are denied access to the data unless they have passed certain authorization procedures. This authorization is often achieved based on cryptographic techniques, whereby information is actually protected by using encryption/decryption keys. However, confidentiality only guarantees that data is not accessible for unauthorized parties; another concept is needed to guarantee that the data is not altered or tampered with; this concept is called the integrity. The latter basically complements confidentiality to ensure that the accessed data is fully protected from unauthorized alterations that may cause a major breach to the accessed system.

2. Authenticity: Authentication has an essential factor to support information security. It actually maintains and controls the access of authorized users to the services they registered for. Authentication can be realized and controlled by cryptographic keys, digital signatures, certificate authority… etc.

3. Performance: The time-consuming tasks of security algorithms have enticed a major interest in the research community. Although securing the different applications is a necessity, security algorithms are known to consume excessive time periods, and this leads to degradation in systems' efficiency. Hence, there should be a balance between the level of protection required and the delays that can be tolerated.

The efficiency of a protected system mainly depends on these aspects and the attention given to them at the design stage. However, since systems and applications have diverse requirements and applications, they will definitely differ in the level of security we can achieve in them. That is, it is not practical to work on fulfilling all of the abovementioned aspects, and certain aspects will get more focus than others. Therefore, we survey in this chapter different security techniques that have been proposed in the literature to serve certain applications. The survey focuses mainly on how the privacy is preserved in the systems provided. After that, we present our new solution that is based on the FHE scheme, and we provide a comparison between our solution and the surveyed proposals.

## 3.2  Overview of Security in Internet Applications

### 3.2.1 Database and Applications

Different security techniques have been devised for relational databases. These techniques can be organized into three categories: encryption, distribution techniques and access control modules. These techniques attempt to satisfy the requirements of database security that include the protection of privacy, data loss avoidance, and the prevention of unauthorized accesses. These requirements have been defined in [SHM10]. In this latter, the data to be protected have been split into three categories:

- Data in use: The data that exists in the memory of the database which is often more important than the data existing in the hard disc.

- Data in motion: The flow of data between both the client and the database server.

- Data at rest: The database stored files.

Shmueli et al. [SHM10] have also categorized attackers into three classes: The intruders, which are external users that gain access to the database, the insiders, which belong to the system and threaten other users, and the malicious administrators, which track the system to extract private information from it. In what follows, we discuss the different security techniques proposed under the abovementioned categories.

### 3.2.1.1 Encryption

Encryption is an effective approach to secure databases. With this approach, database records are encrypted and communications with the database server are made using encrypted queries. Encryption techniques have been widely adopted in the literature as a means to avoid third parties' intervention in the communications between database servers and their clients (see Figure 3-1). Examples of encryption techniques are provided in what follows.

Yin et al. have proposed in [YIN09] an encryption technique based on mathematical transformation. The objective of this technique is to modify the query behavior such that clients interact with the database server in a blind fashion. In the same direction, Popa et al. have presented two mechanisms [POP11]. The first mechanism executes a query over encrypted data to extract the suitable records. The second mechanism is an adjustable technique that customizes encryption scheme according to each query to avoid revealing encryption possibilities.

In another context, Yan and Zhang in [YAN11a] use the homomorphic concept (which we detail later in this thesis) to propose a private homomorphic algorithm that acts on real numbers and supports the different arithmetic operations: addition, multiplication, subtraction and division. The authors have also proposed another mechanism called IOT-MW. This mechanism is responsible for encrypting/decrypting the content and monitoring queries to support communication with encrypted fields.

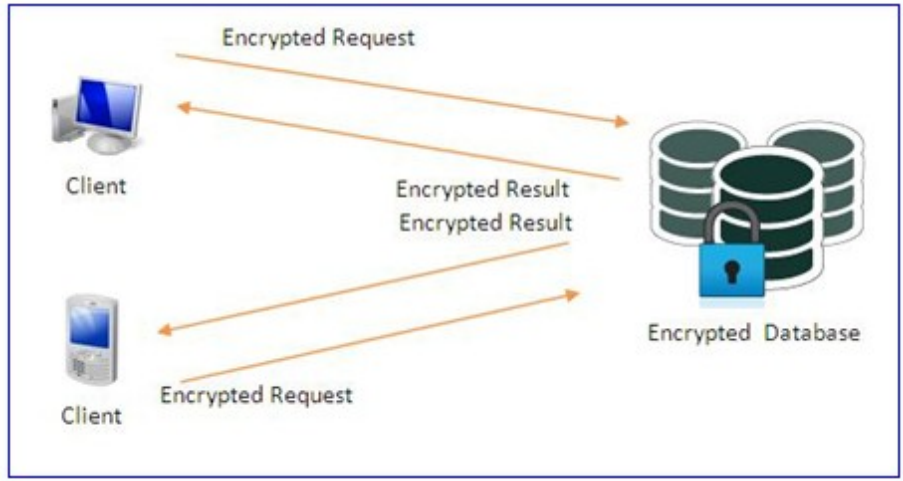

Figure 3-1: Encrypted database architecture without third parties.

Qi-chun has tackled in [QIC09] the problem of executing encrypted queries while considering the time consumed in encrypting the databases. The work in this paper is based on the exponential algorithm and ASCII transformation to reconstruct a new index depending on the query inputs. Then, any record of character type is converted to a larger integer before considering only four bytes of the final converted record. This is meant to achieve a significant decrease in the time needed to perform the query.

The authors in [PAG11] and [RAY11] have followed an approach different than the above mentioned ones. Basically, they depend on third parties to enhance the encryption efficiency. Pagano and Pagano have assumed the existence of a trusted third party that uses a re-routable deterministic encryption technique to link queries with performers [PAG11]. The latter technique is performed without allowing both the performer and the routing module any access to the content of the queries. On the other hand, Raykova et al. have used in [RAY11] a remote synchronizer to store the decryption keys, and maintained a database memory, instead of files stored on hard disc, as part of the proposed architecture.

It is worth mentioning that in enterprises' platforms the outsourced databases are actually utilized in a shared mode where multiple users have access to the same records. In such a context, supporting security becomes challenging and special techniques are needed to guarantee that only authorized users get access to the protected data. Yang et al. have addressed the problem of multi-user settings for encrypted databases [YAN11b]. The paper uses the bilinear maps concept to provide a private keyword search scheme that processes encrypted queries and produces encrypted results from an encrypted database. This is done without disclosing the contents of the queries or the results. The scheme manages also users revoke and denies access without need to a key renewal.

### 3.2.1.2 Distribution

Distributed architectures are often adopted to eliminate complex configurations, and avoid additional middleware that are expensive to repair. Distribution, in general, is realized by distributing processes among several parties and sharing tasks with different modules. This concept can be adopted to build distributed database architectures with an effective security model. Distributed databases aim at storing data in several locations and use multiple performers to manage data according to their sensitivity, see Figure 3-2.

The protection of privacy through employing the distribution concept has received a considerable attention in the research community. The work of Aggarwal et al. in [AGG05] and their extensions proposed by Ganapathy et al. in [GAN11] have presented a distributed architecture that partitions data between two un-trusted servers using a vertical fragmentation.

The system that uses encryption and some obfuscation techniques to support privacy uses a bottom up state algorithm to split queries into sub queries. The latter can be later addressed to different parties without leaving any traces. Also, the system assumes that the different servers are disconnected from each other, and the interrelated sensitive information is not saved in the same server.

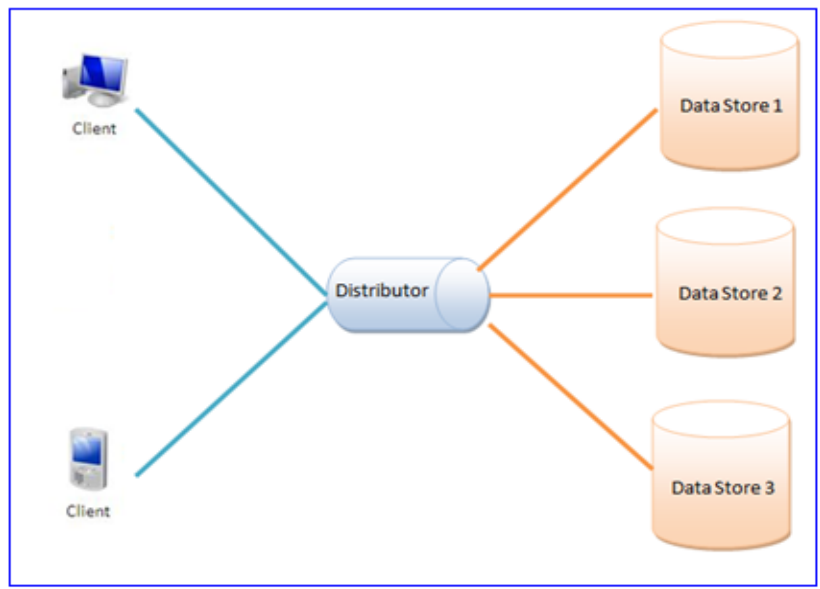

Figure 3-2: Distributed database architecture.

Unlike the previously described work, Yonghong has chosen the horizontal fragmentation and semantic attribute decomposition to partition records over several distributed servers [YON10]. To preserve privacy, a probabilistic anonymity idea is employed when distributing data across servers.

On the other hand, to support the privacy using a distributed architecture and encrypted data, Al thneibat et al. in [THN10] decompose the topology into three components 1) the databases' clients that interact with the database system and submit queries, 2) the service provider that initiates the database structure and provides encryption keys as well as privileges to the third component, and 3) SMM module, which is an interface between the client and the databases where most of the processing is performed. This module is responsible for managing encrypted database access and processing encrypted queries according to the client's authorization.

Indeed, all presented contributions can support the privacy for database systems in different ways. However, there are some theoretical approaches that have not received sufficient attention in this area, such as the Secret Sharing Scheme (SSS). This scheme depends on distributing a secret like encryption keys or sensitive data among a group of users using SSS. This secret is then regrouped and reconstructed by combining all of the parts shared by the participants. Within the same context, Bai et al. in [BAI10] have proposed a secure relational database management system. This system defines a parameter of access K that represents the threshold of privileges. If someone has an access to K or more databases in the distributed architecture, then they are allowed to manipulate the data. Otherwise, the access is denied.

Finally, unlike the approaches that focus on the interaction between the client and the databases, Zhang and Zhao in [ZHA08] have addressed the problem of communication between databases. The authors have considered the queries that aim at finding both the intersection and the difference between several private databases using the two-party concept where each party protects its data and do not want to reveal the content. They have also proposed to protect privacy by controlling query inputs.

### 3.2.1.3 Access control

Access control is the basic technique to protect and manage data access in a database system. This technique assigns privileges to the specific parties that can interact with the topology and then checks authentication parameters to unlock the protected data for processing. The associated architecture is often composed of three modules, the user, the database server and a control panel. The control panel verifies whether the user credentials match the control list or not, and then attributes or denies the access to the server, see Figure 3-3. Such database systems suffer from different types of attacks. The attacker can benefit from failures like privilege abuse, system vulnerabilities and weak audits, to gain unauthorized access to the modules and extract useful information.

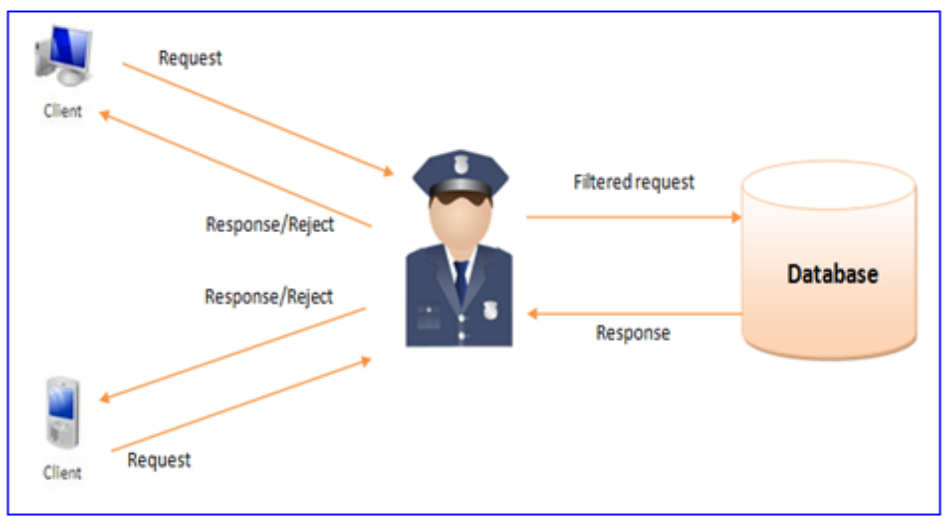

Figure 3-3: Database architecture with control panel.

These attacks can be summarized into three main categories:

1. Brute-force: consists of using default or weak username/passwords to gain access, along with the corresponding privileges, to the databases.

2. SQL Injection: refers to submitting bad SQL queries into systems to change the behavior of the database and trace its responses.

3. Privilege escalation: aims at granting good privileges to a legitimate such that he/she is provided with more authorities than he/she should have.

To face these attacks and protect the privacy, a number of contributions have been proposed. Burtescu paper has presented four axes to deal with these attacks [BUR08]. It advises to ignore the queries that require heavy processing of sensitive records. Then, it suggests hiding the exact value of the returned result, replacing it by an approximate one, and adding fake results in order to confuse the attacker in case its request requires only one answer. Yangqing et al. in [YAN09] have combined an audit module with a rights control mechanism based on the twice login technique to trace requests' behavior and manage access to the database. Furthermore, Ruzhi et al. in [RUZ10] have adopted a third party concept to provide security in the database system. They suggest using a Gateway mechanism that manages all entrances including the database admin. This new module includes four components to ensure an efficient system management. The first component is an authentication model that presents access rules. The second component is a transparent proxy module that protects database configurations. The third component is an attack protection module that detects and prevents different attack techniques. Finally, the fourth component is a connection monitoring module that controls the communication flow between the database and the users. Li et al. in [LI10a] have addressed control access for encrypted databases. They propose a mechanism to authenticate queries for encrypted databases and return encrypted results. Also, they advise to accompany each encrypted result with a proof that helps to reconstruct the result using a chained hashing technique. The authors in [BOU10, AHM11] have considered the special case of web databases access. Ahmad et al. in [AHM11] have considered the multi level access control in web databases and categorized control access into three classes: mandatory, discretionary and role based access controls; whereas Bouchahda et al. have proposed to manage control access based on the application profiles concept [BOU10]. The latter defines and manages a set of SQL queries with specific permission levels varying for one session to another. This approach allows an adjustable control for many profiles.

## 3.2.2 Secure remote execution

Remote execution is a trend by which users, with limited hardware capabilities, can benefit from the powerful resources of a remote server, see Figure 3-4. This can be realized using the cloud computing concept, which provides users with a set of resources and services over the Internet. Cloud computing allows users to upload their binaries and interact with the result, as if these operations were executed locally.

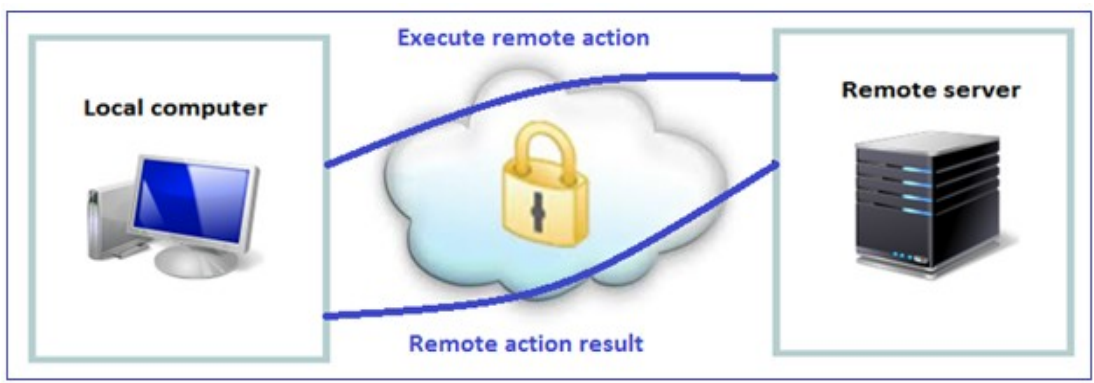

Figure 3-4: Remote execution.

However, from the client side, these operations are actually performed on unknown and untrusted servers. On the other hand, from the server side, the uploaded binaries are not verified, all the time, and may cause real threat. Therefore, several security concerns rise in this context, like having users susceptible to tracking. This necessitates the need to secure the remote execution operation, and the importance of protecting servers from malicious and unknown users. In order to reach these goals, many approaches have been proposed and we organize them into three main categories:

1. **Access control policies**: Program restriction policies are efficient mechanisms that provide servers with the ability to protect themselves against programs at execution. These mechanisms provide some controls that define whether a program is authorized to run. This is besides a set of different execution rules that manage the access process. These rules are useful in identifying the software using some information exported with it such as, the hash information, the certificate authority and the program's zone. The hash information is a cryptographic signature that identifies the software using its associated hash value, the file length and the hash ID. The certificate authority identifies the software publisher, and therefore, helps in making the software trustworthy. The certificate defines a code signing information that points to the owner of that certificate. Finally, the programs' zone refers to the location from where it was uploaded. A number of research studies have focused on enhancing software execution policies such as [PAN00], [KIR02], [SON10], [ABB10], [KIM11] and [ALV10]. In [PAN00], Pandey et al. have provided a general access control policy by defining a declarative access constraint language to restrict access to system components. In [KIR02], Kiriansky et al. have proposed a number of requirements that support security at program execution. At first, access program is restricted depending on its origin. Then, limits are imposed to control access based on the source and the destination as well as the transfer type. Finally, a program is forced to pass through different checkpoints that ensure its trustworthiness. In [SON10], Song et al. have used a logging mechanism that provides access only to allowed components. They have also proposed an architecture that utilizes a key repository, a mandatory access control, and encryption to form an efficient access control method. Abbasi et al. in [ABB10] have focused mainly on web applications, and therefore, have proposed the use of secure web proxies and an extensible access control markup language based on authorization policies. This work has also proposed to secure the communication between the web server and the protected web pages by means of the asynchronous communication protocol. Kim and Kim have introduced a different approach that makes use of virtualization concepts to divide the execution environment into two disconnected domains with different security policies [KIM11]. One domain is dedicated for service requests. It transforms consumer requests into an understandable message, using a multiplexer. Then, the message is submitted to the second domain, which acts as a service provider. Finally, Alves et al. in [ALV10] have improved a runtime platform by modifying the TLS connection to implement a transport component for secure session execution.

2. **Obfuscation**: The obfuscation technique helps in protecting programs against tracking and monitoring as they run remotely. This method is used to transform the source code into another equivalent, but irreversible, form that is difficult to understand. Obfuscation focuses on the protection of the data structure and the control flow of the code. While, the first consists of hiding the source code formatting and the attributes' name, the second consists of obfuscating the control flow of the program that includes program statements, jumps, returns…etc. These controls are modified by either replacing call methods by their body or fake instructions and useless code that sets blind the program behavior. A set of contributions, namely [TOY05], [POP07], [CHE09], [SCH11], [WAN11], [ARM11], and [WEI11], where the authors have addressed programs' obfuscation and have proposed different models to make it hard to reverse engineer these programs. Toyofuku et al. in [TOY05] have proposed the use of random values to hide the control flow of a program. In the proposed technique, an index is assigned to each method and then, during the execution, a random value is generated to decide which method to invoke. This technique makes it hard to follow the execution and protects the program from static analysis. The work in [POP07] makes understanding the program hard by firstly inserting additional statements, and then replacing jumps, method calls and returns by a set of fake instructions that lead to the obfuscated operation. The contribution of [SCH11] is different than the previous ones as it has focused on dynamic analysis. The authors have proposed a mechanism based on diversification concept to disallow running the programs several times and logging their traces. In [WAN11], the authors have proposed the use of fake variables and simple loops that add a huge number of possible execution paths. Armoogum and Caully in [ARM11] have suggested three obfuscation techniques: A variable renaming procedure, a comment removal mechanism, and a control statement insertion that consists of adding useless statements (like if statement, loops, dead code…etc) to confuse the disassembling. In [WEI11], Wei has provided an additional obfuscation technique that consists of transforming functions into a table function that only shows the input and the output data. Finally, Chen et al. in [CHE09] have proposed a binary obfuscation technique based on the taint tags and the opaque predicates.

3. **Cryptography**: Encryption has often been used to protect sensitive data in several areas against malicious and unauthorized users. It consists of employing and managing a set of encryption/decryption keys that lock/unlock a specific piece of data. The same concept has also been used to encrypt some software modules in order to obfuscate their behavior and secure the whole architecture. The contributions [PAT07], [SON10], [ABB10], [BIS07], [WEI11], and [BRE11] are key examples of such architectures. In [PAT07], Patel et al. have proposed an architecture where the program information are encrypted using a secure key and stored in a dedicated processor. Whereas the authors of [SON10], [ABB10] and [BIS07] have chosen encryption techniques that aim at protecting web page contents, data and software by storing decryption keys and allowing only authorized users to access their content. In the same direction, the work presented in [WEI11] has advocated the use of RSA encryption and the impossibility of factorization of prime numbers to control the computational complexity. On the other hand, Brenner et al. have discussed in [BRE11] a different concept and have focused on computations over encrypted data using HES. This work has proposed a scheme that designs encrypted circuits using an integer representation, and allows a performer to operate over them without being aware of their content.

### 3.2.3 Location Based Services and Mobility Solutions

In the recent decades we have noticed that both the growth of the smart phone industry and the rise of connectivity have resulted in the emergence of a diverse set of products and services. These products and services have paved the way for the development of new promising business models. Among these technologies, Location-Based Services (LBSs), see Figure 3-5, have attracted the utmost attention in the research community. In general, we shall mention that these services rely on the development of wireless networks, the innovation of positioning services, and the availability of information. These three elements are mainly used to determine the current position of a user, and therefore, locate the closest businesses or services (banks, restaurants, universities…etc) around that user. The user can enquire about that information by communicating wirelessly with an LBS server.

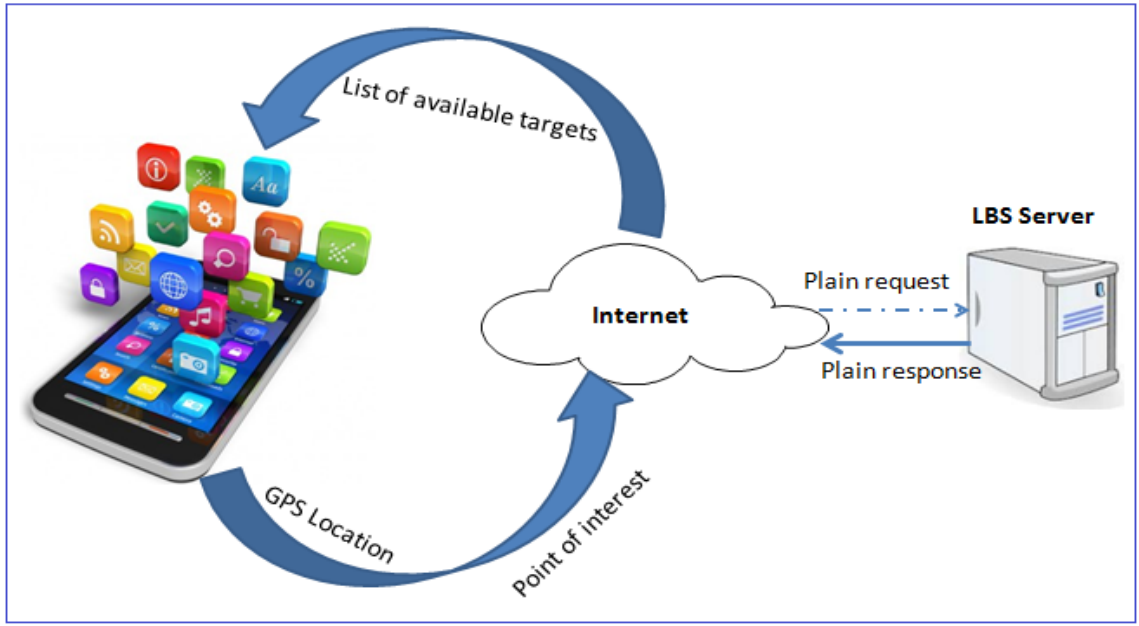

Figure 3-5: Basic LBS architecture.

However, since this technology deals with sensitive information (like the current position of the user and his/her daily habits), providing a secure architecture that allows the abovementioned elements to interact anonymously between each other becomes an essential task. Therefore, several approaches have been devised to resolve this problem. These approaches provide various secure models for the LBS technology, and aim at protecting the privacy of the enquirer and the location whereabouts. The contributions that aim at protecting the enquirer's information and hiding their real identity tend to prevent LBS servers from collecting information to identify the users. The approaches that we are referring to employ anonymity techniques such as [XU07], [XU08], [GRU03], [GED05], and [MOK06] or pseudonym like [JOR07] and [ZEI08]. They all aim at obfuscating the real use and mislead the service providers. The anonymity techniques work on protecting user identity from its neighbors precisely. This means that when a client enquires about a specific location, the anonymizer module replaces the query by a box that combines the user location with those of its neighbors, and sends it instead of the query that was originally requested. The anonymizer then distributes the generated response to the appropriate user while making use of the real identity. Within the same principle, pseudonym is another technique by which users protect their identities. This technique uses fake identifiers instead of the real ones and delegates the mapping between them to a location intermediary. Jorns et al. in [JOR07] have designed a system architecture for using safety LBS while relying on pseudonym transaction. The latter consists of generating unique pseudonyms for each transaction such that no fake identifier can be reused again within subsequent operations. Using a different approach, the work presented in [ZEI08] relies on random numbers and strings to generate strong pseudonyms that are used as a basis for an LBS authentication mechanism.

Within the same context of enforcing the privacy and security of the user's location, several techniques have been developed. The main concepts that have been utilized in these techniques are relying on K neighbors, sending a cloaking region instead of the true location, and initiating several fake requests. The first technique, which is K-anonymity, depends on hiding the location information among the K-1 surrounding locations. Then, instead of working with the real location, the combination of all neighbors' locations is employed (see [REI79], [GED04], [GED05], and [MOK06]). This concept has been extended to provide another technique called the cloaked region. This technique was first proposed by [KAL06]. It depends on using the dimension of a particular region, the size of which depends on the specified security level, and protecting the location privacy by sending requests to the whole region instead of one location. On the other hand, Kido et al. [KID05] and You et al. [YOU07] have proposed a different method to reach the same goal. Their technique relies on a third party component that adds to the original request many fake queries, with different fake identities, to prevent LBS from revealing the true information. Furthermore, Argadna et al. [ARD11] have proposed an obfuscation technique that transforms the location measurements by changing both the radius and the center, and then sends the obfuscated location to LBS. Wightman et al. in [WIG11] have focused on three pre-existing techniques: the randomization, K-anonymity and clocking region, and then proposed new extensions. Basically, the authors have introduced three techniques: N-Rand, N-Mix and N-Dispersion. While the randomization technique consists of changing the center of an area by another random point to allow users obfuscate their real location, the N-Rand enhances this concept by introducing another parameter that defines the suitable distance of obfuscation. The N-Mix technique relies on randomly generated locations instead of adopting the basic concept of K-anonymous, which relies on pre existing users' location. Finally, the N-Dispersion technique enhances the concept of dispersion, which consists of de-centering the circular area, by performing n trials while still looking for the new displacement. A different but simple technique has been introduced by He et al. in [He11]. It works on cheating on the sent location by modifying the GPS API and sending fake locations instead. The authors in [JI11] have relied on K-anonymity to define different cells in order to hide the path a user follows from source to destination, whereas the contribution of Pingley et al. has developed query perturbation-based scheme that relies on generating several fake queries that take into account the query context [PIN11]. Finally, Chen and J. Pang have proposed in [CHE12] a different scheme that relies on some metrics to design the most secure region for a specific user profile.

## 3.2.4 Ad-Hoc Wireless Networks

Ad-hoc wireless networking is the revolutionary approach that allows all the devices that belong to the same communication range, and support the necessary wireless equipment, to directly exchange information in a peer-to-peer mode, see Figure 3-6.

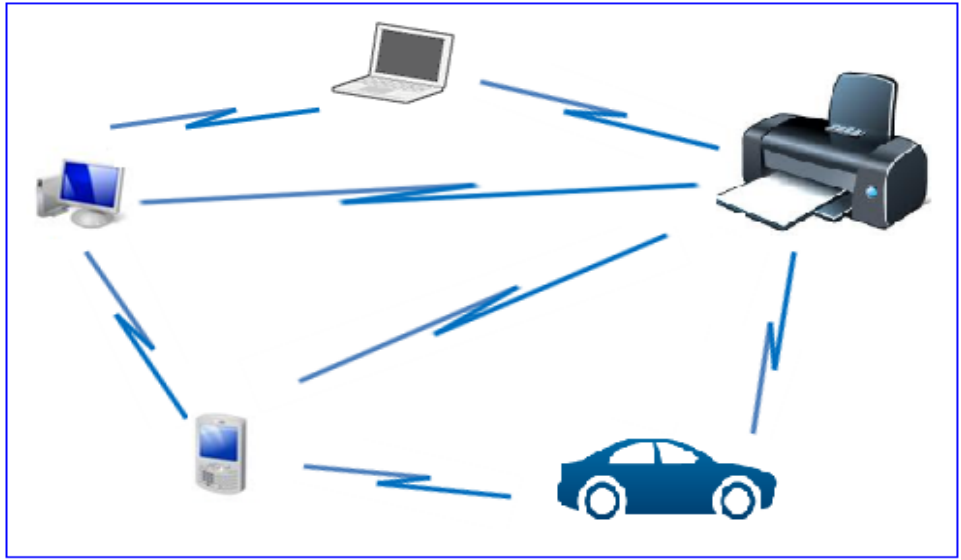

Figure 3-6: Ad-hoc wireless environment.

These devices, called nodes, join and leave the network dynamically and collaborate in a benevolent manner by forwarding packets to other neighboring nodes. However, many challenges arise against the efficiency of the ad hoc network's architecture because of the nodes' availability problem, and the fact that this technology does not rely on pre-existing topologies. The latter factor has received a special attention in the literature, especially that is has a direct impact on routing protocols. In general, the routing concept relies on the voluntary cooperation of various nodes that are able to forward packets from a given source to a specific destination. Given the importance of routing in the network, guaranteeing a secured topology is essential to preserve the privacy. This is because malicious and selfish nodes, if available, may mislead nodes' collaboration and make packet delivery probabilistic. In order to avoid this problem and ensure a secure exchange of information, several mechanisms have been proposed. The most common mechanism in this context is the trust-based methodology. This mechanism enforces the fact that each node evaluates its neighbors and assigns them trust values that are continuously updated based on their degree of cooperation. Hence, trust values are used to define a trust route. Within the same context, Gera et al. in [GER10] have based their work on an effective combination of both a trust value approach and a multipath route technique. From one side, the trust value approach depends on collaborations with only faithful nodes. On the other side, the multipath route technique chooses several paths to reach the destination and forwards the packet over them in order to increase the delivery rate. Parvin et al. in [PAR10] have proposed the use of a given threshold to decide whether to collaborate with a specific node or not. Basically, the proposal uses different parameters, like the requesting information, the type of data, and the exchange history, to calculate a value that is associated with each node. If a node's associated value is greater than the set threshold, then it is allowed to be part of the routing process. El-Bendary et al. in [BEN11] has focused on the direct diffusion routing algorithm that mainly relies on sending low frequency requests and acts according to the feedback (positive or negative). This algorithm has enabled the authors to design an authenticated acknowledgment-based protocol. This protocol inherits many features from this technique, like the propagation phase and the authentication stage that validates the ACK message, this latter is used to confirm the delivery. On the other hand B-Tebibel in [BOU11] has proposed to protect the routing protocol using a hashed, one-time passwords protocol that consists of checking each hop while constructing the path. Within the same context and using a different approach, the technique that has been presented in [LI11] depends on an identity-based routing protocol that uses digital signatures to ensure that the collaborating nodes are credible. This protocol also checks the remaining energy in each of these nodes to confirm their readiness to participate in the route. Similarly, the contributions in [MAH11] and [ZHA12] have used the energy levels as well as the past behavior of the nodes to decide which path is more secure. The authors in [FAG11] have used a different approach that relies on the information from the application layer to design a cross-layer approach, which monitors the network and put away the packets from the danger zone, and a routing protocol. This approach consists of broadcasting a route discovery action by sending a probe message, and then locates the area where malicious nodes exist in order to successfully route the packets away from them. Finally, Abumansoor and Boukerche in [ABU11] have proposed to use the node's location information, mobility, and stability to decide whether it can be trusted as part of the route or not.

### 3.2.5 Video On-Demand Services

In the recent decades, the growth of Internet infrastructure and the improvement of various streaming protocols have resulted in the emergence of a wide portfolio of interesting services. Among these services, Video On-Demand (VOD) has attracted the utmost attention. VOD enables users to retrieve and have full access of a large set of videos at their convenience, and this has contributed to the widespread popularity of VOD, see Figure 3-7.

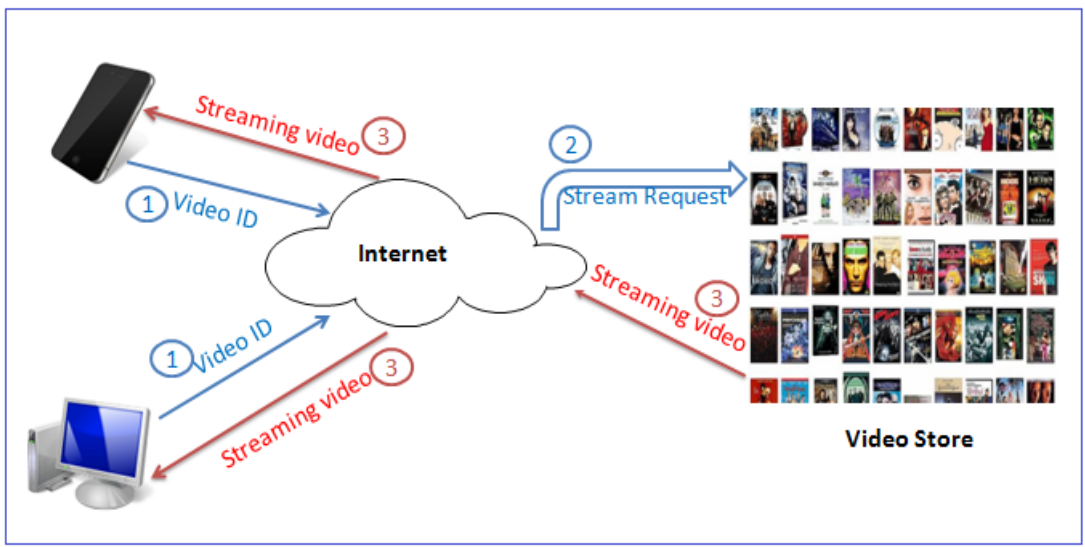

Figure 3-7: Basic VOD architecture.

However, this useful service faces major security challenges, especially related to the confidentiality of service providers and the privacy of end users. Service providers seek to protect their own video store by granting access only for authorized clients. On the other side, clients require that service providers do not monitor or track the different services they purchase. In order to achieve these levels of protection, several secure protocols have been proposed and studied. Most of the suggested protocols have adopted encryption techniques that obfuscate videos content and enforce a secure transfer.

Lee et al. have used the digital watermarking technique to protect video content and trace illegal distribution [LEE10]. This technique primarily encrypts the video, by applying different XOR operations, and then transmits it to the appropriate destination which is encrypted using a suitable key as well. From the client's perspective, the video is watermarked using the appropriate user ID in order to deny any illegal distribution. The work presented in [ROY11] has proposed a different approach that makes use of both compression and joint encryption to protect the video content. This means that the encryption scheme enforces the use of two different encryption keys instead of one. The first key, named the temporal key, is used to encrypt every single frame in the video. The first temporal key, associated to the first frame, is generated from the video key created by the video provider. Every other temporal key is hashed from its previous one associated to the previous frame. Because every frame is divided into several slices, the scheme makes use of a second key, called the spatial key, to encrypt each slice. This powerful technique aims at protecting the system from any frame loss incidents. The authors in [LI10b] have combined a hierarchical encryption strategy with XOR operations to provide a secure encryption scheme for the video content, and allow only authorized nodes to decrypt the data. The work in [VAR11] has relied on Galois field polynomials to provide a robust encryption algorithm. This algorithm encrypts the intra frame by using secret sharing techniques and provides safe entities with the requested access. To achieve the same goal of encrypting the video content, Yeung et al. have used in [YEU11] the 8x8 block technique, while the authors in [DUB11], [JAN11], and [SWA12] have chosen partially encryption techniques to increase the performance of their systems. On one hand, Dubois et al. have relied on a selective encryption technique that partially encrypts the content of the video [DUB11]. Their approach firstly analyzes each macro block separately, using some metrics and measures, and then decides whether it will be encrypted or not without interfering with the level of security. On the other hand, the work in [JAN11] has relied on the expected quality, the available information about the content, and the multimedia scene to make excessive use of the Huffman code compression and partially encrypt the video content. Similarly, Swaminathan, and Mitra have partially encrypted in [SWA12] the video contents based on the entropy and the sensibility of the samples. Due to the importance of the privacy issue, Varalakshmi et al. have come up in [VAR12] with a different approach that combines both the permutation code algorithm and dynamic keys in order to successfully protect confidential video transmission. In this approach, the permutation table is calculated by means of DCT coefficient and video motion measures.

It is undeniable that the security of Internet applications, services, and routing protocols can be achieved by employing different techniques. However, the most common approach followed by researchers is to adopt a third party concept, which is considered a major hole in the architecture because it may be a single point of attack. As a result, a system that ensures a direct secure communications between all parties can be more successful, and therefore, we advocate the use of HES as they are well aligned with our objectives.

## 3.3 Homomorphic Encryption Schemes

Homomorphic encryption allows performers to compute correct operations over encrypted values without being aware of their content. This possibility is due to the fact that the encryption circuit is defined as a group homomorphism, which preserves operations in the group. Group homomorphism makes the computing over encrypted or plain values of the same effect. This flexibility resolves security issues in a variety of applications that delegate sensitive processing to un-trusted third parties.

In this section, we overview the history of homomorphic encryption and describe its mechanisms and characteristics. We also highlight different applications where the involvement of HES is beneficial.

### 3.3.1 History

The existence of an efficient homomorphic encryption has been standing an open question for a long time. The first model has been proposed in 1978 by Rivest, Adleman, and Dertouzos [RIV78]. They have used the exponentiation and large integers to design an additive and multiplicative homomorphic circuit called the RSA. RSA uses the fact that it is difficult to factor large prime numbers to define a strong public key cryptosystem. This key is used to encrypt data and perform simple operations over them. Many other works (like [GOL82], [ELG84], [PAI99], [DAM01], and [GAL02]) have followed the same concept in order to provide more efficient homomorphic schemes. The GM scheme [GOL82] proposed by Goldwasser and Micali is a probabilistic asymmetric public key encryption system that produces a ciphertext of size bigger than its associated input plaintext and then performs additions modulo 2 over that ciphertext. ElGamal in [ELG84] has used the Diffie-Hellman key exchange concept to define another probabilistic asymmetric key encryption algorithm based on exponentiations. Following these attempts, Paillier has proposed in [PAI99] the first secure and efficient additive scheme; it relies on operations in the ring of integers modulo $P^2$ where P is the product of two large primes. Thereafter, this interesting scheme has been the subject of other extensions (see [GAL02] and [DAM01]) that aimed at generalizing its basic concept. These extensions have used either the context of elliptic curves or computing in the rings of integers modulo $P^{s+1}$, where Paillier's scheme is the case with s = 1. Despite the obvious progress achieved in performing computations over encrypted values, the cited schemes are considered only semi-homomorphic since they support either additions or multiplications, but not both operations at the same time. Therefore, providing a scheme that supports both additions and multiplications has remained an open challenge until 2005, when Boneh et al. published their scheme in [BON05]. Although the latter could only support a single multiplication and many additions, it has paved the way for other attempts to provide robust schemes, such as the FHE that has been introduced by Gentry [GEN09a].

## 3.3.2 The Somewhat Homomorphic Encryption Scheme

The first FHE scheme that has been proposed by Gentry [GEN09a] was a form of encryption that supports performing arbitrary additions and multiplications, at the same time, on encrypted values. It uses the polynomial form $c = pk * q + 2 * r + m$ to define a probabilistic public key cryptosystem, where $m$ is the bit value to encrypt as $c$, $r$ and $q$ are two random integers, and pk is the public key with the requirement that 2*$r$ is smaller than $pk$ /2. Then, the decryption is correctly retrieved by carrying out two modulo operations such that $m = c \bmod sk \bmod 2$, where $sk$ is the correspondent secret key. In more details, the proposed scheme consists of four main modules:

1) **KeyGen** ($\lambda$): This module outputs two random values $pk$ and $sk$, which are the public and secret keys, respectively. These outputs are based on the parameter of security $\lambda$ that specifies the length of encryption keys and the encrypted value.

2) **Encrypt** ($pk, m$): This module encrypts (transforms) a bit value {0, 1} into a big integer at the order of $\lambda^n$-bit number that has the same parity as the original bit value, where $n \in \mathbb{N}$.

3) **Decrypt** ($sk, c$): This module decrypts the input ciphertext $c$ based on the appropriate secret key $sk$.

4) **Evaluate** ($pk, C$, *): This module presents the ciphertext result of the performed circuit $C$ over the encrypted values.

The goal of this scheme, which is called a Somewhat Homomorphic Encryption Scheme (SHES), is to provide conditioned number of operations over protected data. This condition relies mainly on the fact that the ciphertext $c$ mod $pk$ (called the noise of the scheme) should be smaller than $pk/2$. However, this noise value, which doubles after each addition and squares after each multiplication, exceeds the threshold $pk/2$ after a finite number of operations, and the correct decryption will not be guaranteed anymore. Therefore, Gentry has proposed a bootstrapping procedure to remove the noise and provide augmented number of algebraic computations.

### 3.3.3 Bootstrapping and the Fully Homomorphic Scheme

The key idea behind bootstrapping is to re-encrypt a bounded ciphertext to refresh its noise value and then support more computations. The bootstrapping encrypts a fired ciphertext $c$ into a new ciphertext $c^+$, which is a double encryption of the plaintext, with a fresh noise. Furthermore, all homomorphic properties are applied on that new ciphertext, which is an encrypted form of the old one. Thereafter, after many computations and before the encrypted result exceeds the threshold, one can perfectly decrypt that result to retrieve another ciphertext that has the same parity as the plaintext.

The bootstrapping technique uses a public key to re-encrypt every bit in the cipher text and produce a new clean cipher text with smaller noise. Then, it uses an encrypted private key to remove the inner layer of the encryption and extract the original cipher text. It is apparent that the original plaintext is simply a double decryption of the bootstrapped values.

By adopting a bootstrapping technique, The SHES produces a FHES that supports unlimited number of both additions and multiplications. It is worth mentioning, however, that bootstrapping is a time consuming task since re-encryption is usually performed over big integer values.

## 3.4 Conclusion

In this Chapter we have highlighted a number of techniques that aim at securing applications, services, and routing protocols. However, we have seen that most of these techniques still suffer from the leak of security and need to be enhanced. Therefore, we have discussed the importance of HES and how they can protect the privacy of users. In the next chapters, we use HES to propose novel models with efficient architectures for sensitive applications.

# Chapter 4

# Secure Database System using Homomorphic Encryption Schemes

## 4.1  Introduction

Cloud computing is an attractive solution that can provide low cost storage and processing capabilities for government agencies, hospitals, and small and medium enterprises. It has the advantage of reducing the IT costs and providing more services for the requesting parties through making specialized software and computing resources available. However, there are major concerns that should be considered by any organization migrating to cloud computing. The confidentiality of information as well as the liability for incidents affecting the infrastructure arise as two important examples in this context. Indeed, cloud computing poses several data protection risks for the cloud's clients and providers. For example, the cloud's client may not be aware of the practices according to which the cloud's provider processes the stored data. Therefore, the cloud's client cannot guarantee that the data are processed (for example, altered or deleted) in a legal and accepted manner.

All of the above mentioned issues can be resolved if the data in the cloud are stored and processed in encrypted and blind form. The latter is possible if the encryption scheme can support addition and multiplication of the encrypted data. A cryptosystem which supports both addition and multiplication (referred to as FHES) can be effective data protection, and enables the construction of programs that receive encrypted input and produce encrypted output. Since such programs do not decrypt the input, they can be run by an un-trusted party without revealing their data and internal states. Such programs will have great practical implications in the outsourcing of private computations, especially in the context of cloud computing and remote databases. In theory, the data in such secure systems could be encrypted by the client, and then sent to the cloud's provider for storage or processing. Only the client holds the decryption keys necessary to read the data. Indeed, this model of computing can preserve the confidentiality and integrity of the data while delegating the storage and processing to an un-trusted third party.

In this chapter, we present a novel technique to execute SQL statements over encrypted data. We develop a secure database system that processes these queries. The parameters of SQL queries are encrypted by the requester and sent to the server for processing. The latter performs the requested operation over an encrypted database and returns an encrypted result to the client. The advantage of this system is that the database server knows neither the content nor the position of the records affected by the query.

Despite the fact that this type of processing may increase the amount of computing time, the benefits associated with it are worth the processing overhead.

The remainder of this chapter is organized as follows. In Section 4.2, we review the literature for the work related to private information retrieval (PIR) approaches. Section 4.3 provides a formal description of our secured SQL statements approach. Section 4.4 presents a homomorphic cryptosystem that we use to build a prototype system. Section 4.5 presents an implementation of a secure relational database system. In Section 4.6 we provide performance analysis of the proposed secure database system. Finally, Section 4.7 concludes our work and provides future research directions.

## 4.2  Private information retrieval

Chow et al. [CHO09] discussed the importance of cloud computing, and how this technology can be enticing due to its flexibility and cost-efficiency. The authors pointed out that the adoption of such technology is still below ambition. Some users are still concerned about the security of these clouds. Even those who started using the technology, they only utilize it with their less sensitive data. The limited usage of cloud computing is mainly due to the lack of control over the communicated data. The authors highlight that people require explicit guarantees that their data will be protected under well-defined policies and mechanisms. However, no technical security solutions were proposed to back-up their information centric model where data can self defend itself in a hostile or an un-trusted environment.

The PIR approach, introduced by Chor et al. in [CHO98], achieves the retrieval of an $i^{th}$ bit in a block without revealing information about the bit retrieved or about the request for the bit itself. This approach has been widely used as a basis for several tools, and has supported various distributed applications. However, the approach requires more improvements and the work with it is still in progress, both at the security of the communication channel level and the hidden client identity level.

Raykova et al. [RAY09] extended the PIR approach by proposing a secure anonymous search system. The system employs keyword search such that only authorized clients have access to their blocks. This system is capable of mapping the database content to the appropriate client, thus guaranteeing the privacy of the data and the query. The ultimate target of Raykova's system is to ignore the identity of the client while protecting the database from malicious enquirers.

Shang et al. [SHA10] tackled the problem of protecting the database itself. The problem is studied through monitoring the amount of data disclosed by a PIR protocol during a single run. The information attained from the monitoring process is used to understand how a malicious enquirer can conduct attacks to retrieve excessive amount of data from the server.

PIR has also been used to develop authentication systems. Nakamura et al. [NAK09] constructed a system with three components, a enquirer that initiates requests, an authentication-server that processes these requests, and a database that returns the appropriate data in response to the request. This system ensures the security of data and the anonymous communication between the enquirer and the database. Yinan and Cao [YIN09] used the PIR approach to propose a system that controls the access to the database. According to this system, the privacy of data is enforced by enabling each authorizer to give or deny access to his/her own data with a hierarchical authorization access right scheme.

Among the most important criteria in PIR protocol are the communication cost and the amount of data sent back to the enquirer. The trivial solution of the PIR protocol is to send back the entire database to the client. However, this solution is expensive, even for a simple request that results in retrieving two matching records. Other approaches proposed to retrieve only the requested data, by using replicated databases that are stored at multiple servers. In this case, the request is forwarded to all servers. With this approach, although we deal with multiple replicated databases, the privacy is better protected. However, this approach is still complicated and may result in extended processing and communication times. Gentry et al. [GEN05] proposed a scheme to retrieve a bit or a block from a database with a constant communication rate. Melchor et al. [MEL08] proposed a scheme that reaches the available data with a reasonable communication cost while achieving lower computational cost compared to other PIR protocols.

## 4.3 Secured SQL operations

In this section, we develop a secure database system that processes SQL queries over encrypted data. As shown in Figure 4-1, parameters of the queries are encrypted by the client and sent to the server for processing. The latter performs the requested operation and returns encrypted results to the client. We describe below the circuit of a simple SQL SELECT query:

*SELECT * from T where c=v*

where the value *v* is in encrypted form. The trivial solution to securely perform this statement is to send back to the client the entire database, but this solution suffers from complexity and scalability issues. Instead, we propose a methodology to implement the SELECT circuit at the server side, while preserving the confidentiality and the privacy of the request.

The processing of the SELECT query is divided into three sub-circuits. Firstly, we calculate the following index for each record *R* in the table *T*:

$$\forall R \in T \ I_R = \prod_{i=0}^{size-1} (1 \oplus c_i \oplus v_i)$$

where *size* is the number of bits in column *c*; $c_i$ and $v_i$ are the $i^{th}$ bits of column *c* and search criteria *v*, respectively. $I_R$ is a one bit value that is equal to 1 if *v* matches the value of column *c*, 0 otherwise. $\oplus$ is a XOR operation.

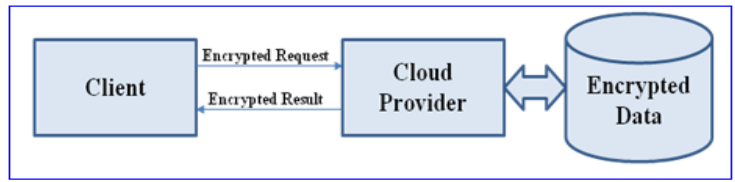

Figure 4-1: Secure Data Retrieval.

Next, we identify the $n^{th}$ record that matches the selection criteria. For that purpose, we consider $\eta = \varepsilon_{pk}(n)$ to be the encryption of $n$ under public key $pk$.

For each record $R$ we calculate the following sum:

$$\forall R \in T: S_R = \sum_{i \leq R} I_R$$

We calculate a second index $I'_R$:

$$\forall R \in T \; I'_R = I_R \times \prod_{i=0}^{size-1} (1 \oplus \eta_i \oplus S_{R,i})$$

$I'_R$ is equal to 1 if the record $R$ is the $n^{th}$ record that matches the selection criteria, 0 otherwise. Then, we multiply every bit of each record $R$ in table T by the corresponding value $I'_R$ to construct a new record $R'$.

$$\forall R \in T: R' = I'_R \times R$$

This latter operation forms a new table $T'$ that is related to the original table $T$ with the following characteristics:

$$\begin{cases} R' = R \; if \; R \; is \; the \; n^{th} \; \text{record matching the criteria} \\ \quad\quad R' = 0 \;\; \text{otherewise} \end{cases}$$

Finally, by adding all records of table $T'$, we retrieve the $n^{th}$ record $R_s$ that matches the selection criteria:

$$R_s = \sum_{R' \in T'} R'$$

If no record matches the selection criteria, a record containing zeros will be returned to the requester.

It is worth noting that all calculations are performed over encrypted data. The server does a blind processing to retrieve the $n^{th}$ record that matches the selection criteria. It neither has access to the content of the retrieved record nor to its position within the table.

With slight modifications to the select circuit, most of SQL operations can be supported by our proposed secure database system. For example, to implement the UPDATE operation, one can simply implement the following circuit:

$$\forall R \in T: \ R^{'} = \overline{I_R} * R + I_R * U$$

where the record $U$ is the new value to update the record $R$ matching the criteria of the query.

Similarly, to delete a record from table $T$, one can replace its content by zero. The DELETE operation can be implemented by the following circuit:

$$\forall R \in T: \ R^{'} = \overline{I_R} * R$$

## 4.4 Homomorphic encryption scheme

As mentioned in the previous chapter, Gentry proposed a FHES that enables to perform an arbitrary number of arithmetic operations (i.e. addition and multiplication) on encrypted data. The components of a boostrappable encryption scheme, already described in sections 3.3.2 and 3.3.3, are detailed below.

In what follows we consider the security parameters: N = $\lambda$, P = $\lambda^2$, and Q = $\lambda^5$.

## 4.4.1 Key generation

The private key *sk* is a random $\lambda^2$-bit odd number p. The public key consists of a list of integers that are the "encryptions of zero" using the encryption scheme with the secret key *sk* as a public key.

Generate a set $\vec{y} = \{y_1, \ldots, y_\beta\}$ of rational numbers in [0,2[ such that there is a sparse subset $S \in \{1, \ldots, \beta\}$ of size $\propto$ with $\sum_{i \in S} y_i \approx \frac{1}{p} \bmod 2$, and with the condition: $0 < \propto < \beta$.

Set sk* to be the sparse subset S, encoded as a vector $s \in \{0,1\}^\beta$ with hamming weight $\propto$.

Set pk* $\leftarrow (pk, \overrightarrow{y})$ to be the public key.

## 4.4.2 Encryption (pk*,m)

Set $m'$ to be a random $\lambda$-bit number such that m and m' have the same parity:

$$m' = m\%2$$

Then compute *c* as:

$$c \leftarrow m' + pq$$

where *q* is a random $\lambda^5$-bit number. Then the ciphertext *c* is post-processed to produce a vector $\vec{z} = \{z_1, \ldots, z_\beta\}$, defined by:

$$z_i \leftarrow c. y_i \bmod 2$$

The output ciphertext *c** consists of *c* and $\vec{z} = \{z_1, \ldots, z_\beta\}$.

### 4.4.3 Decryption (sk*, c*)

$$m \leftarrow \text{LSB}(c) \oplus \text{LSB}(\sum_i s_i . z_i)$$

where LBS means the least significant bit.

### 4.4.4 Arithmetic operations

Addition and multiplication can be performed on clear text by simply adding and multiplying the ciphertexts, respectively.

$$\varepsilon_{pk}(m_1 * m_2) = \varepsilon_{pk}(m_1) * \varepsilon_{pk}(m_2)$$

$$\varepsilon_{pk}(m_1 + m_2) = \varepsilon_{pk}(m_1) + \varepsilon_{pk}(m_2)$$

The output ciphertext *c\** consists of *c* together with the result of post-processing the resulting ciphertext with $\vec{y}$.

### 4.4.5 Bootstrapping the Encryption Scheme

The encryption scheme described above is referred to as a SHE because it works only if value of the ratio *c%p* (noise of the encryption) is smaller than *p/2*. After a finite number of arithmetic operations, the noise exceeds the *p/2* threshold and the decryption does not work anymore.

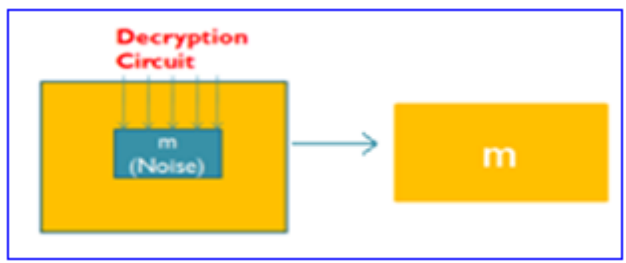

Figure 4-2: Removing noise from original ciphertext (bootstrapping).

Gentry developed a novel method to remove the noise in the ciphertext [GEN09a]. He proposed to recrypt the ciphertext $c$ to remove the noise. Since the scheme is homomorphic, one can encrypt the ciphertext $c$ into a new ciphertext $c^+$ (the plaintext is encrypted twice), and by using the homomorphic properties of the scheme, one can decrypt the inner layer of encryption to obtain a ciphertext c2 with a lower value of noise.

As illustrated in Figure 4-2, a bit $m$ is encrypted with public key $pk$ to produce the ciphertext $c_1$. After a finite number of arithmetic operations, the noise associated with the ciphertext $c_1$ reaches a level that does not permit any additional arithmetic operation. To remove the noise, the bootstrapping technique consists of recrypting the bit $m$. Every bit of ciphertext $c_1$ is encrypted with the public key $pk$. The output is ciphertext $\bar{c}$ that doubly encrypts bit $m$. The decryption circuit is applied to remove the inner layer of encryption. This latter operation requires the knowledge of the private $sk$. Therefore, the private key is encrypted with public key $pk$; and then shared with the server. Since the encryption scheme is homomorphic, the decryption can be performed on the doubly encrypted ciphertext to remove the inner layer. The recryption produces a new ciphertext $c_2$ with a value of noise that has an upper bound according to the proof in [GEN10a].

Indeed, the bootstrapping is a time consuming task. To technically show that, we measured the time required to perform the product of two n-bits numbers in encrypted form using the fully homomorphic cryptosystem (with bootstrapping).

Figure 4-3 shows the amount of time, in seconds, required to compute this homomorphic multiplication circuit .

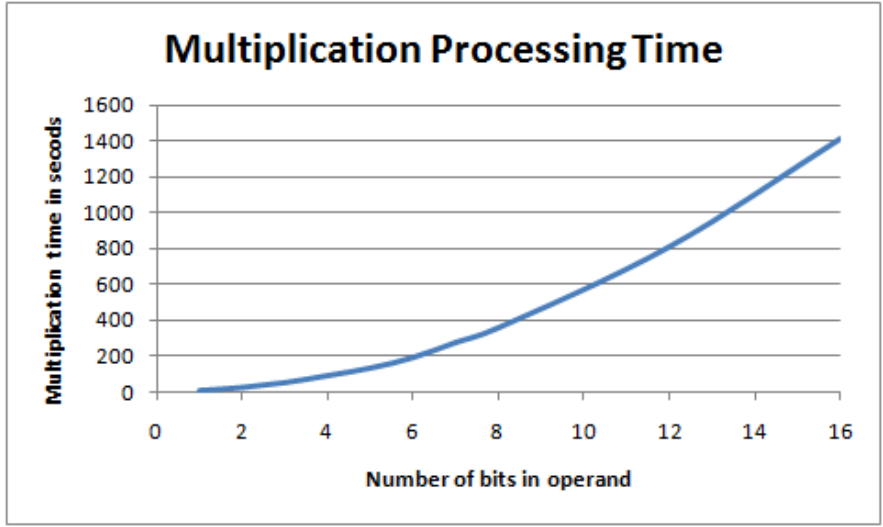

Figure 4-3: Processing time required to perform the product of two n-bits integers.

As we can see in Figure 4-3, the circuit takes 23 minutes to compute the product of two 16-bit integers. This latency is mainly due to the recrypt function.

Despite the amount of time needed and required while employing the bootstrapping technique, performing fully secure systems in a blind fashion becomes possible and all these performance issues could be fixed.

## 4.5  Implementation

In our implementation, we aim at proving that it is possible to perform SQL queries over an encrypted database. For example, the user can specify a search criterion through a database. Then, the client software encrypts the parameters of the query, corresponding to the search criterion, and sends it to the appropriate server. The server retrieves the requested record (blind processing) from the database and returns it to the client. The client software decrypts the record and displays it to the user.

We built a simple medical application containing 10 patients' records. In Figure 4-4, we show the result of the SELECT query. This is how the result appears in a screenshot of the client side of our built application. The application supports the following SQL operations:

- SELECT with wildcard characters (*, ?) and relational operators (< >).
- UPDATE with wildcard characters (*, ?) and relational operators (< >).
- DELETE with wildcard characters (*, ?) and relational operators (< >).
- Statistical operations like COUNT and AVG.

It is worth mentioning that the implementation of the medical application was built using a somewhat homomorphic scheme. This is due to performance issues as it is impractical to perform our tests using the fully homomorphic cryptosystem. We chose the security parameter in such a way to support all the SQL operations with no need to employ the bootstrapping $\lambda=3$ technique. We discuss the performance of our system in the next section.

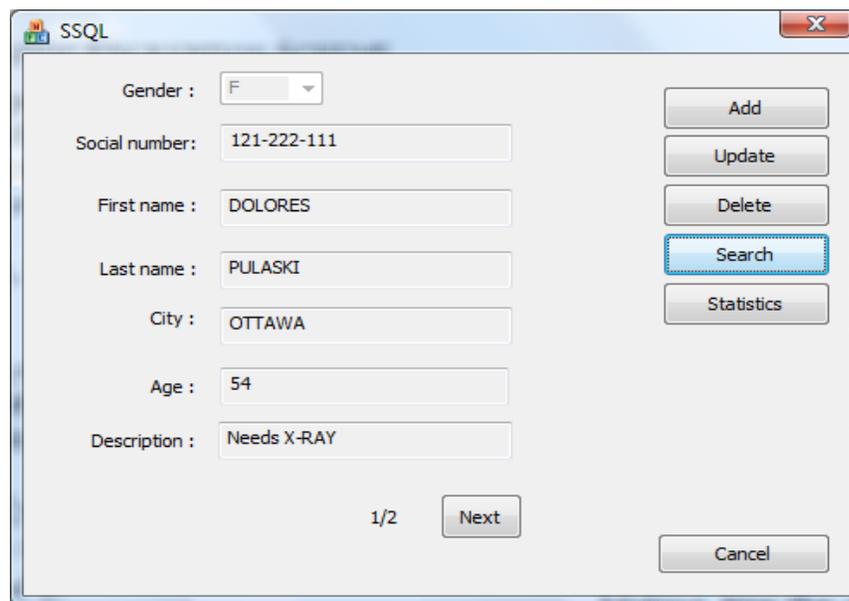

Figure 4-4: Client side of the application.

## 4.6    Performance analysis

Table 4.1 lists the number of arithmetic operations required to execute some basic SQL statements over an encrypted database of 10 records. From this table we can see that processing data in encrypted form creates a substantial computation overhead.

|  | Add. & Mult. | Add. | Mult. |
|---|---|---|---|
| SELECT | 619839 | 309892 | 309947 |
| UPDATE | 67595 | 25355 | 42240 |
| DELETE | 28171 | 5643 | 22528 |

Table 4.1: Number of arithmetic operations

Taking into account that with such configuration, we need 2 seconds and 1 second to perform $10^4$ Mult and $10^5$ Add, respectively. That means, the select circuit takes more than one minute to perform a search request in that database .

Towards that end, we used a computer machine with 1.7 GHz processor and 3GB of RAM memory. Microsoft Visual studio 2008 C++ has been used to develop the application.

The implementation of the system proves that the execution of SQL statements over encrypted data is feasible. However, the time required to execute these statements is very high and therefore is not suitable for real-time transactions that involve a large database (i.e. several terabytes database). This drawback is mainly due to the FHES. In fact, there might be more efficient techniques to optimize the implementation, that is, one could perform recryption only when it is necessary, since the noise value can be bounded; however, we do believe that a more practical homomorphic cryptosystem is yet to be developed.

## 4.7  Conclusions and future directions

The concept of processing encrypted data is promising to revolutionize traditional computing. Indeed, this concept has many direct applications in cloud computing environments, banking, electronic voting and many other applications.

In this chapter we developed the first secure database system based on a FHES. We presented the circuits to implement SQL statements over encrypted data. We built a prototype of a database system where data is stored and processed in encrypted form. The database server can execute most of the SQL statements in a blind fashion, that is, it returns the results without any knowledge of the content or the position of the records extracted/affected. We conducted performance analysis to measure the time needed to execute a simple query on the database. We found that the current technology is not sufficiently mature yet as it is time-consuming. Indeed, the encryption schemes proposed by Gentry et al. in [DIJ10][GEN10a][GEN09a] are very impractical. According to our measurements, the time needed to perform simple calculations is substantial. We believe that there still is a great opportunity for researchers to develop more efficient HES.

As future work, we are planning to work on the optimization of the efficiency of the system. Processing can be parallelized in order to take advantage of multiple processors executing the encrypted requests. We will also investigate how to reduce the number of recryptions needed. Indeed, since the noise value can be bounded, decryption should be necessary only when the ciphertext cannot support an additional arithmetic operation.

In the current system, the server does know the operation that was performed (SELECT, UPDATE, etc.). If we can encrypt the SQL circuits, the system will preserve the confidentiality of the data and operations performed on these data. We believe that this new system can be the foundation of a highly secure cloud computing environment. The next chapter tackles this problem by proposing a system working over encrypted processes.

# Chapter 5

# Encrypted Processes for Oblivious Data Retrieval

## 5.1 Introduction

The cloud computing technology is a distributed system by which multiple remote servers and applications are made available over the Internet to provide various capabilities. The latter include data storage, utilizing remote services, and handling complex computations that require powerful hardware resources. The main purpose of cloud computing is to extend the resources of each computer using the Internet. The cloud operates in an open environment and therefore several precautions should be taken to preserve the privacy of its clients. Although the cloud employs a set of protection techniques, these may not be effective in all situations, especially when dealing with sensitive data. In fact, clients with sensitive data require strong security measures before they can trust any cloud service. As a result, cloud's provider must have the ability to (a) guarantee the privacy of clients' data and (b) protect the security and privacy of clients operations in the cloud against the threat of de-compilation.

Providing these abilities entice more clients to trust the cloud computing technology, however, has been a challenge as they require delegating computations without delegating access. Therefore, the need for a mechanism to allow the cloud to communicate with the client in a blind manner has been an important quest, where the cloud is able to execute clients' queries without having access to their contents or behavior. Such a mechanism can be achieved by utilizing the most recent version of HESs [GEN10a], which is the subject of this chapter.

HES allow performing computations over encrypted data, based on addition and multiplication, without decryption. With these schemes, a client is able to publish only encrypted data and ask un-trusted parties (a cloud computing provider) to perform computations on his/her behalf. Using homomorphic encryption, an un-trusted cloud provider can perform some operations on the data without the need to decrypt the data. The final result of these computations is equivalent to these computations being performed on the clear data. This means that homomorphic encryption allows clients to preserve the privacy of their data, as only encrypted data will be available to the service provider. However, this technique is time-consuming as it executes a large number of operations to perform the requested queries, which is a major drawback. Nevertheless, the benefits of homomorphic encryption motivated research efforts to develop techniques that require lower delay overhead.

In this chapter, we introduce novel homomorphic encryption circuits that are an extension of our previous chapter. In this latter, we proposed a set of homomorphic circuits that perform simple database queries over a database of encrypted data, without allowing the owner of that database to know the targeted records. In this chapter we use HES to develop a generic circuit that performs any database queries, instead of using n-circuits for n-different operations. Besides, we propose a mechanism to encrypt the circuit itself. The latter is achieved through circuit obfuscation. An obfuscated program is a circuit that hides the program's original functionality while maintaining the same behavior. This technique is used to protect the internal state of a process from de-compilers and malicious attacks. As a result, the parties performing computations on behalf of the client will neither have information about the data being manipulated nor the arithmetic operations (*addition* or *multiplication*) that are being executed.

The rest of this chapter is organized as follows. In Section 5.2, we review the research contributions in the area circuit obfuscation. In Section 5.3, we describe our technique to encrypt processes for oblivious data retrieval. In Section 5.4, we provide an implementation based on our scheme for a secure database system. Finally, Section 5.5 concludes our work.

## 5.2  Related Work

Circuit obfuscation has attracted research efforts in the last decade. However, mixed findings have been reported in the literature. In some cases, researchers have reported negative results about obfuscation [BOA01 and GOL05], while other researchers have demonstrated the opportunity to build an absolutely oblivious program that keeps the privacy of operations [KAI10, HAS09, SHA06, JIE07, GOL05, and LIN03]. In this section we discuss the findings of all of these studies.

Boaz et al [BOA01] studied the possibilities to conceive a general obfuscator that preserves the same functionality as the original program, and whether such an approach can be widely used to protect software from attacks. The conclusion of their study was absolutely negative. They proved theoretically that such an obfuscator cannot be achieved by studying a set of functions that are not obfuscatable, such as one-way functions. However, the authors clearly indicated that homomorphic encryption techniques could be a promising track to construct a robust obfuscator system.

Goldwasser and Kalai [GOL05] extended the theoretical work of [BOA01] and confirmed the limits of these obfuscators' robustness. They studied more functions that cannot be obfuscated, such as filter functions.

In [GEN10a], Gentry demonstrated the most recent homomorphic encryption theories that paved the way for new promising techniques to build strong obfuscators. Kai-Min et al. [KAI10] used the theories presented in [GEN10a]. Their main idea is to check the loyalty of the remote party where the computation was delegated. However, checking if the remote agent gave back the right result does not guarantee the trustiness of the party. Therefore, it will be interesting to prevent a server from knowing what data is being manipulated.

The authors in [HAS09], [SHA06], [JIE07], and [POP07] described various techniques that protect mobile agents while they are under full control. Hashmi and Brooke [HAS09] proposed a paradigm to protect mobile agents at runtime from static and dynamic attacks, while this program turned in un-trusted hosts. The paradigm consists of detecting if an agent meets an attack from malicious software using various parameters, such as the time calculated for an agent to finish its task in a specific environment. Shah et al [SHA06] proposed obfuscation techniques that act on the byte codes of java based mobile agents. Their techniques deny any attempt of reverse engineering by a deviant class file to extract the original source code. Jiehong et al [JIE07] used two techniques, namely, obfuscated control flow and time checking, to prevent malicious programs from knowing a mobile agent's state. The first technique consists of substituting a set of instructions by a more complicated equivalent one, while the second technique calculates the execution duration to check if the node has performed more computations than needed. Popov et al [POP07] combined the obfuscated control flow, mentioned above, with another technique that inserts some fake instructions to make it difficult to find the beginning of each instruction at runtime.

Linn and Debray [LIN03] used the technique of obfuscating source code to make a statistically hard disassembling program.

It is worth mentioning that all of the above techniques do not depend on encryption to fully protect the targeted program; they just produce an obfuscated form that makes it hard to understand the source code of a program. Therefore, the internal state and program's behavior are not fully protected. In this chapter, we focus on the problem of performing private operations over encrypted data. This approach can effectively hide the internal state or the functionality of the targeted program. That is, we show that with our approach we are to secure the system against various malicious attacks.

## 5.3  Secure Data and queries behavior

Let us consider the following HES: $\varepsilon_{pk}(m) = c = m + 2 * r + pk * q$; where c is the cipher text of bit m; $pk$ is the public key; and $r$ and $q$ are two random integers [GEN10a]. The decryption scheme is as follows: $m = (c \bmod sk) \bmod 2$ where $sk$ is the secret key. By carefully choosing the size of the keys $pk$ and $sk$ and the random values $r$ and $q$, the encryption scheme is proved to be semantically secure [GEN10a]. This encryption scheme uses logical expressions that are implemented by a set of logical gates that form a system to achieve a specific need. It depends on the *XOR* and *AND* logical gates to implement the addition (*add*) and multiplication (*mult*) of ciphertexts. However, the fact that some systems consist of gates interrelated in a specific manner makes them very weak against de-compilation. As a result, most systems, especially those involving applications of databases, can be easily monitored. For example, an attacker can detect what operations a client is requesting by noticing the gates that are in use and the information disclosed about their interconnection. It is essential to hide that low level information in order to provide higher level of protection for the clients' privacy. To accomplish this protection, we propose to encrypt the information about the gates in use. Towards that, we deploy a black box that encrypts both *add* and *mult* gates to avoid all drawbacks resulting from disclosed schemes. The encryption of these gates propagates to the encryption of the whole circuit, forcing a malicious host to execute requests in a blind manner. This novel technique, implementing a new gate -named the Star gate, gives a requester the ability to preserve the privacy of his/her operations, and prevents the server executing the operations from tracing his/her steps. In Figure 5-1, we show a schematic of the Star gate, which performs the following logical relationship through three encrypted bits X, Y, and S:

$$f(S, X, Y) = (S \times X \times Y) \oplus [\overline{S} \times (X \oplus Y)]$$

The Star gate takes as input three encrypted parameters: The two bits $X = \varepsilon_{pk}(x)$ and $Y = \varepsilon_{pk}(y)$, which are two operands to calculate, and $S = \varepsilon_{sk}(s)$, which is used to configure the output to be either $X \times Y$ or $X \oplus Y$. In fact, the decision of which operator the Star gate will generate as an output has been delegated to an encrypted item to maintain the privacy of the process. Therefore, we should not worry about the credibility of the executer, as they conduct the orders in complete ignorance. In a simple way, if we need to perform a safety operation *mult* or *add*, we just have to execute the Star by giving $S = \varepsilon_{sk}(1)$ for the first operation and $S = \varepsilon_{sk}(0)$ for the second, as shown in the two next relations:

$$\varepsilon_{pk}(x * y) = \varepsilon_{pk}(1) * \varepsilon_{pk}(x * y) + \overline{\varepsilon_{pk}(1)} * \varepsilon_{pk}(x + y)$$

$$\varepsilon_{pk}(x + y) = \varepsilon_{pk}(0) * \varepsilon_{pk}(x * y) + \overline{\varepsilon_{pk}(0)} * \varepsilon_{pk}(x + y)$$

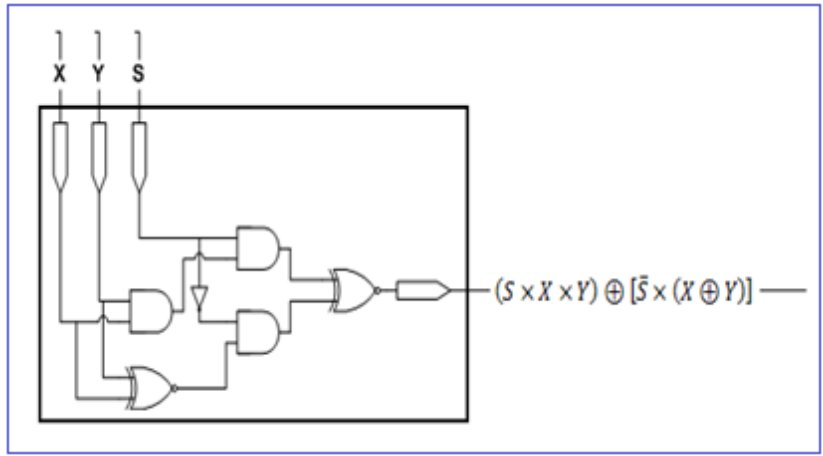

Figure 5-1: Star gate.

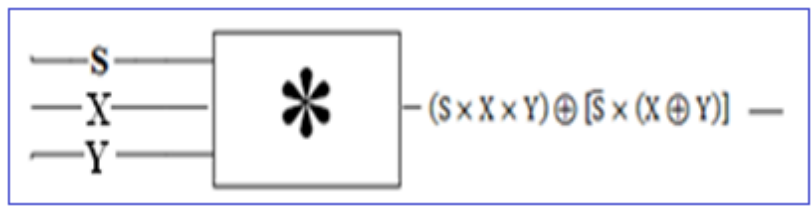

Figure 5-2: Simplified representation of the star gate.

Indeed, the representation of our schemas becomes complex as shown in Figure 5-1, since we replace each *mult* and *add* by a Star, which is not a simple gate. Therefore, to avoid this drawback we will adopt a new abstraction. In Figure 5-2, we provide a simplified representation of the Star.

As systems based on logical expressions are widely accepted in cryptography, the encryption of logic gates described above can be a strong way to make such systems highly secure. For the rest of this section, we will use this encryption of gates to transform a partially secure system, which executes database queries, into a fully secure one without affecting the system's functionality (hereafter called Secure-Executer).

In the previous chapter, we have developed a secure system that executes simple database queries over encrypted data to prevent the server from knowing which records are being manipulated. We proved that our approach could preserve privacy. However, the server could still know which operations were being executed. Such information may be used by malicious database owners to retrieve clients' orientation relying on the nature of operations they perform. Therefore, we needed to develop a common tunnel to perform all operations, instead of creating n tunnels for n different operations, where the circuits are encrypted using the Star gate. The combination of these two techniques creates a new mechanism that performs database queries in a secure way, where the database owner has no knowledge of what the requested operations (*SELECT*, *UPDATE*…etc) are.

The new system operates in three main steps, namely, localizing, identifying, and creating results. Basically, each sub-circuit initializes its flags and switchers from the client request as well as the request criterion *v*. The latter contains information about what the requester is searching the database for. The flags are used to configure the Star gate to perform an *add* or a *mult*, while the switchers are used to lead the circuit and draw the circuit path.

The localizing step is an essential one when performing a request. This step focuses on finding each record *R* that matches the request in a database *T* by comparing the criterion *v* to each column *c* in the database's records. This information is used in all the other sub-circuits that follow. Regardless of the operation sent by the client, the localizing step must be able to locate suitable records by turning over encrypted data. For this purpose, we designed a localizer schema that is made up of several branches, as shown in Figure 5-3. Only the requester has the right to decide the appropriate branch to use since each branch is associated with a specific operation.

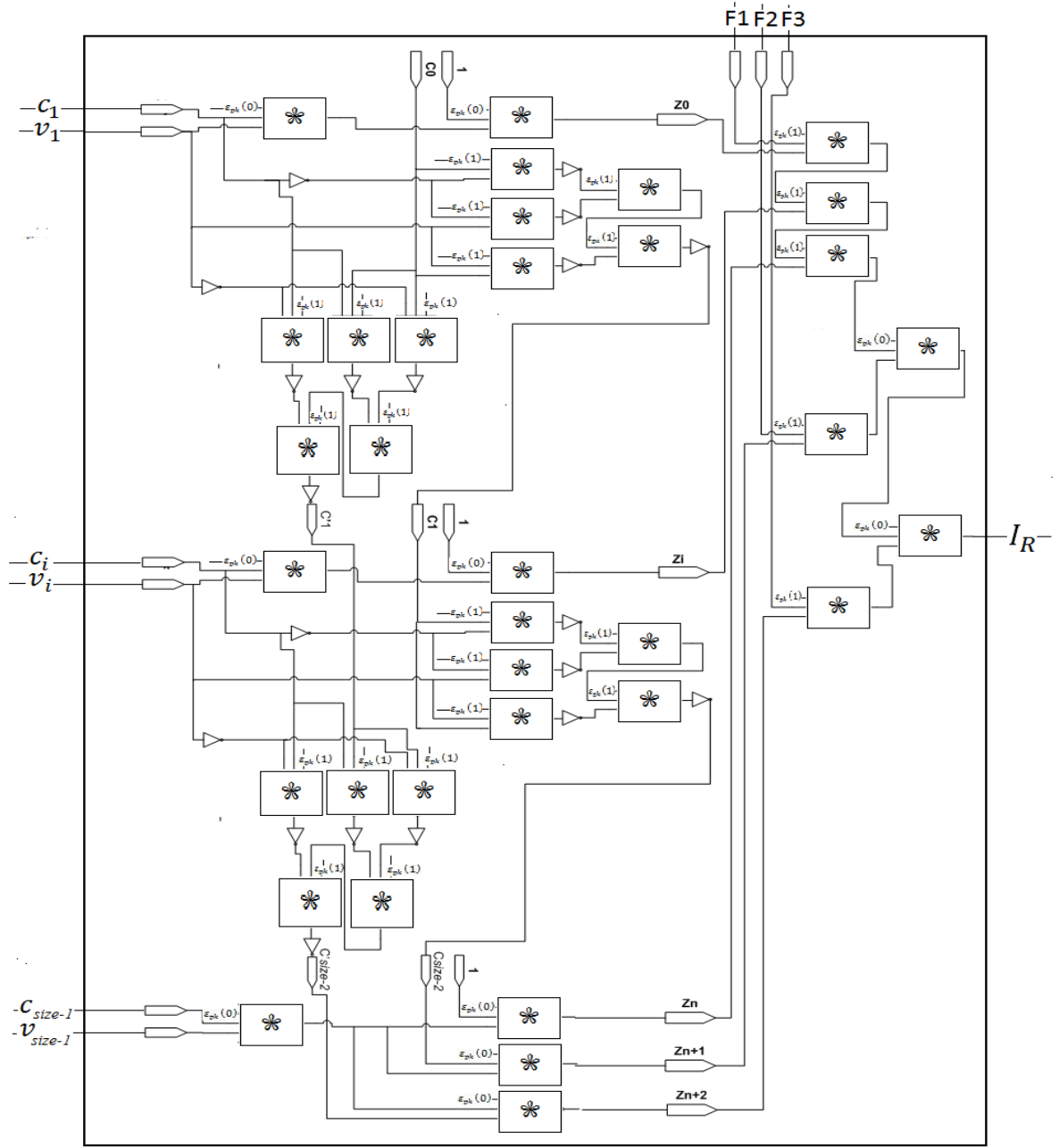

Figure 5-3: The localizer schema.

Whatever decision the client makes, the server will be ignorant of it, as the bit received by the server is encrypted. In other words, if the client chooses the encrypted bit value F = $\varepsilon_{pk}(1)$ to be associated with a branch, the specified branch is taken into consideration and F = $\varepsilon_{pk}(0)$ will be ignored.

The description discussed above can be formalized as: $\forall R \in T$:

$$I_R = \left[\!\!\left[ F1 \times \prod_{i=0}^{size-1} (1 \oplus c_i \oplus v_i) \right]\!\!\right] \oplus \left[\!\!\left[ F2 \times C_{R,size-1} \right]\!\!\right] \oplus \left[\!\!\left[ F3 \times C'_{R,size-1} \right]\!\!\right]$$

With:

$$C_{R,i} = \begin{cases} 0, & i = 0 \\ 1 \oplus \left[\!\!\left[ \overline{\overline{c_i \times v_i}} \times \overline{\overline{c_i} \times C_{R,i-1}} \times \overline{v_i \times C_{R,i-1}} \right]\!\!\right], & i \geq 1 \end{cases}$$

And:

$$C'_{R,i} = \begin{cases} 0, & i = 0 \\ 1 \oplus \left[\!\!\left[ \overline{\overline{v_i \times c_i}} \times \overline{\overline{v_i} \times C_{R,i-1}} \times \overline{c_i \times C_{R,i-1}} \right]\!\!\right], & i \geq 1 \end{cases}$$

Where size is the number of bits in column c, $c_i$ is the $i^{th}$ bit of column c, and $v_i$ is the $i^{th}$ bit of the request criterion $v$. F1, F2, and F3 are the three switchers sent by the client to decide which branch to use and which ones to ignore. The result $I_R$ is a one bit value that is equal to 1 if v matches the value of column c, 0 otherwise. In the next sub-circuit, we identify the $n^{th}$ record that matches the selection criteria. For that purpose, we consider $\eta = B(n)$ to be the binary form of the index n represented by $size$ bits.

For each record R we calculate the following sum:

$$\forall R \in T: S_R = \sum_{i \leq R} I_R$$

Then, we calculate an index $I'_R$ :

$$\forall R \in T\ \ I'_R = I_R \times \prod_{i=0}^{size-1} (1 \oplus \eta_i \oplus S_{R,i})$$

$I'_R$ is equal to 1 if the record R is the $n^{th}$ record that matches the selection criteria, 0 otherwise.

At the final step, the creator applies the request to the T as follows: $\forall R \in T$

$$R' = \left[\!\left[(F4 \times I'_R) \oplus \overline{I'_R}\right]\!\right] \times R \oplus \overline{\left[\!\left[(F4 \times I'_R) \oplus \overline{I'_R}\right]\!\right]} \times U$$

Where F4 is an encrypted flag that indicates the read/write operation's state and U are the new values to put into the database. These values are equal to 0 if the client does not request an update to his/her records. This generic formula also allows hiding the operating state of R/W. Thus, in consultation and deletion cases, the true operation is hidden by adding a misleading update that does not change the content. Our system can also guarantee a highly secure system that can preserve all transactions' state and content.

This latter sub-circuit returns to the requester a result from his/her query. The result contains records from the SELECT request, or an indication that the records have been deleted (after a DELETE request) or updated (after an UPDATE request). This result is a table T', which is built according to the following:

$$\forall (R, R') \in (T, T') : \ R' = \begin{cases} R \text{ if R is matching the criteria} \\ \quad 0 \ \text{otherewise} \end{cases}$$

Finally, by adding all the records of table $T'$, we can retrieve the $n^{th}$ record $R_s$

$$R_s = \sum_{R' \in T'} R'$$

If no record matches the selection criteria, a record containing zeros will be returned to the requester.

## 5.4 Implementation and tests

In this section we present a prototype, named the Encrypted Database Queries Executer (EDQE), which we use to prove that it is possible to perform queries through encrypted processes. As shown in Figure 5-4, our system consists of three modules, the Requester, Encryption/Decryption, and the Performer. Once a client initiates a new database access, the Requester retrieves the request and prepares a datagram, as shown in Figure 5-5, that contains the search criterion, flags that suit the query and, the new data, in case the client wants to perform an UPDATE. Then, the datagram is sent to the Encryption/Decryption module to be encrypted in. The encryption is done to prevent the Performer from knowing what the client is requesting. This performer uses the encrypted datagram to search in an encrypted database where each record is $15 \times 10^4$ bit length. Once it finishes performing the request, it sends the result back to the Encryption/Decryption module for decryption. Finally, the latter module terminates the circuit by sending back the plain result to the Requester to display it.

Database queries are executed in the same general circuit controlled by the input flags. So, the fact that these flags are in an encrypted format, as shown in Figure 5-5, will prevent any third party from tracing our system. This makes our system highly secure against traceability and attacks.

The prototype supports the following database operations:

- SELECT with wildcard characters (*, ?) and relational operators (< >).
- UPDATE with wildcard characters (*, ?) and relational operators (< >).
- DELETE with wildcard characters (*, ?) and relational operators (< >).
- Statistical operations like COUNT and AVG.

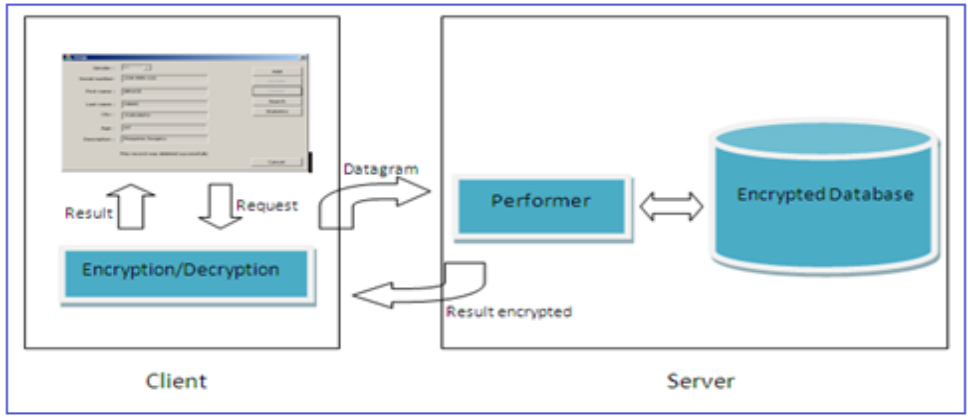

Figure 5-4: Circuit of performing requests.

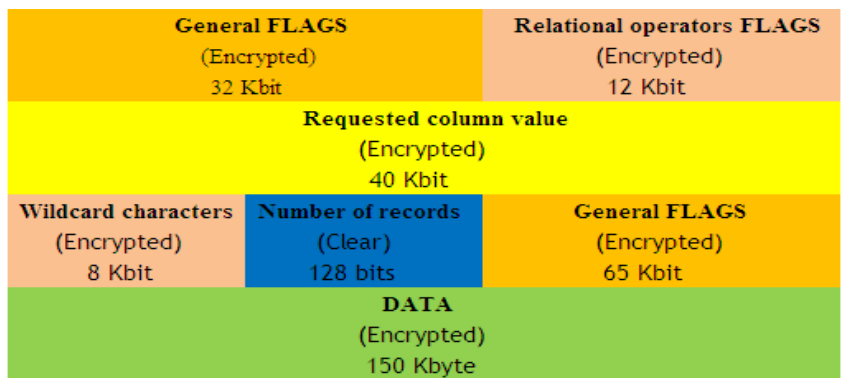

Figure 5-5: Datagram.

In Figure 5-6, we show the result of the DELETE query.

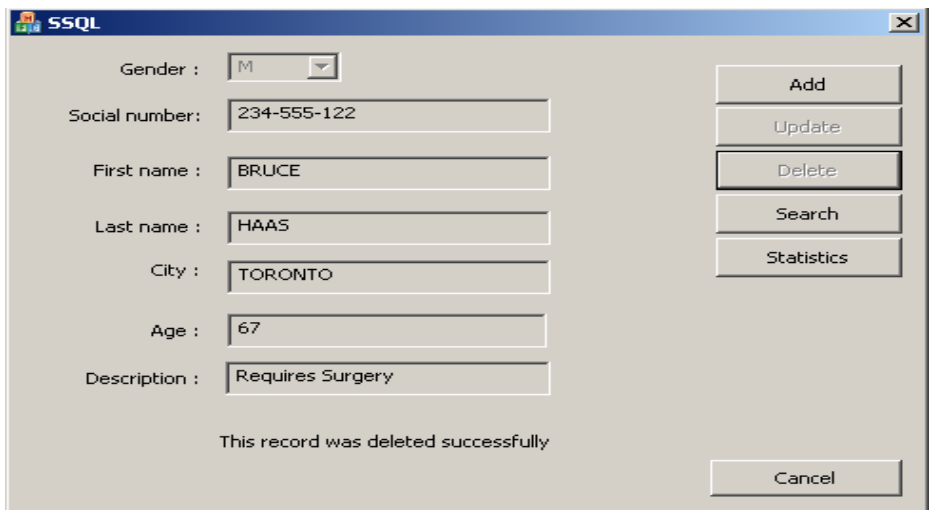

Figure 5-6: Client side of the application.

It is worth noting that all database statements require the same number of arithmetic operations. These statements differ only in the value of the encrypted flags sent by the client.

Table 5.1 lists the number of arithmetic operations required to execute the general circuit over an encrypted database of 10 records.

|  | Add & Mult | Add | Mult |
|---|---|---|---|
| **General circuit** | 819892 | 409892 | 410000 |

Table 5.1: Number of arithmetic operations

We note that we are using a simplified version of the FHES, called SHES. This technique enables us to execute encrypted processes within reduced time compared to the fully one, since it consumes less operations.

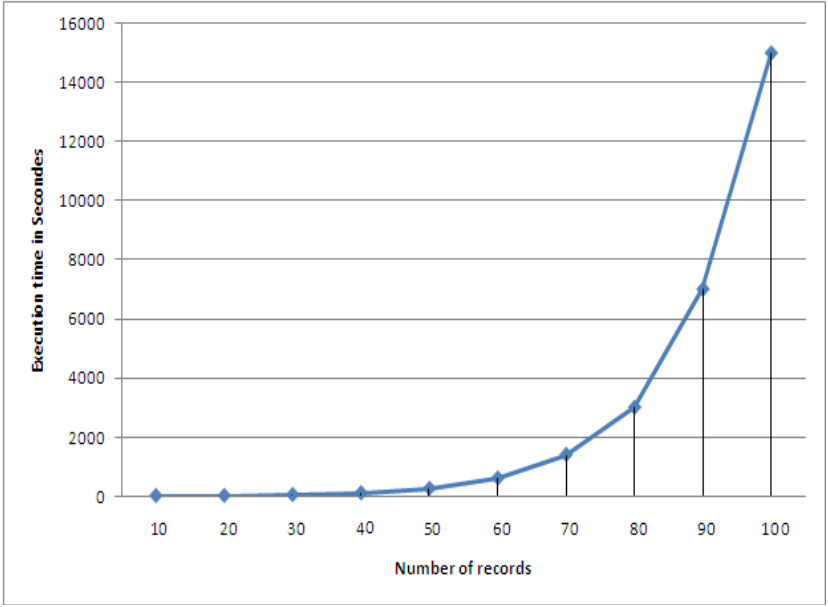

Figure 5-7: Execution time.

Figure 5-7 illustrates the performance in terms of the processing time consumed for different number of records. These results show that our system still suffers in performance, but it can be considered as a first step towards a practical system that preserves the privacy of a process executed in a hostile environment.

## 5.5 Conclusion

Homomorphic encryption is a promising approach to strengthen the security measures of computing systems that manipulate sensitive data. This technique allows clients to rely on the services offered by remote servers, even though the integrity of these servers may be questionable.

In this chapter, we have presented two novel techniques that use homomorphic encryption to preserve the confidentiality of clients' data when storing this data at remote servers. In the first technique, we developed a global homomorphic circuit that processes all of the database queries that a hosting server can execute on behalf of the client, in order to hide the operation's nature . In the second technique, we strengthened the concept by encrypting the operation itself in such a way that the servers performing the operation can learn nothing about the requests of the client. Combined together, the techniques provide a complete confidentiality of both the data manipulated and the nature of queries being performed.

In the next chapter, we customize the models developed in Chapter4 and Chapter5 to fully secure an interesting area, which is a mobility solution. We also enhance the approach by considering the noise value and giving related statistics.

# Chapter 6

# Privacy Preserving Scheme for Location-Based Services

## 6.1  Introduction

The growth of smart phones and mobile devices in both software and hardware capabilities have resulted in the emergence of a set of new products and internet services that guarantee new promising business models. Location-Based Services (LBSs) have attracted the utmost importance in this regard. These services rely on the Global Positioning System (GPS) or Network-Based Positioning, which are mainly used to determine the current position of a user, in order to define his/her location relatively to a business or a service (banks, restaurants, universities…etc). The user can enquire about that information by communicating wirelessly with an LBS server. The server uses the signal emitted by the user to locate him/her using Real-time Locating Systems (RTLS) [CLA09]. Once the coordinates of the user are determined, the server responds with a list of all services surrounding the user's position.

LBSs have actually attracted the research and development community. However, LBSs suffer from a major security pitfall in terms of violating users' privacy. In other words, as the LBS server gains knowledge of the users' coordinates, this information can be manipulated by the server itself or by any malicious party to trace the movements of the users. Thereby, instead of using such a mechanism to facilitate lifestyle, it can easily turn over into an efficient tracking tool. This problematic urged the research community to find a secure way to use LBSs without disclosing users' private information.

Strong protection for users' information can be attained if the server is made capable of retrieving location-related information without being aware of the user's position or the point of interests he/she is requesting. It is challenging to achieve the latter target as the server needs to at least know this search criterion to retrieve the requested information. In this chapter, we tackle this problem by using encryption schemes to retrieve data without violating the privacy of the users.

The remainder of this chapter is organized as follows. In Section 6.2 we review the related work that aimed at securing location-based services. Section 6.3 provides a detailed description of the circuits that makes it possible to respond requests in a blind fashion. Section 6.4 presents our prototype and the evaluation of its performance. Finally, Section 6.5 concludes our work and provides future research directions.

## 6.2 Related work

There are a number of approaches in the literature to solve the problem of privacy protection with location based services, including:

- Cloaking.

- Generation of dummies.

- PIR.

Gruteser and Grunwald [GRU03] and Chow et al. [CHO06] have based their approaches on K-anonymity [GED05, MOK06, REI79, and GED04]. The latter concept relies on hiding the user's location among K-1 neighbors. The main idea behind this concept is to send a box of locations instead of only the true one, whereby the probability to guess the user's location is always less than 1/K. All techniques relying on K-anonymity use a middleware (the anonymizer) [GRU03, CHO06, GED05, MOK06, REI79, and GED04]. This anonymizer is a third party responsible for creating a Cloaking Region (CR), which contains the true user's location, as well as K-1 other neighbors. With such a technique, a typical scenario can be a user trying to localize the nearest bank. The user sends his/her requests (including his/her credentials) to the anonymizer through a wireless network. Thereafter, the anonymizer, which keeps the locations of all current users, authenticates the requester first and chooses a set of K-1 neighbors to create a CR that can be sent instead of the user's position. This way, the risk of violating the user's privacy is reduced by making it difficult to locate the position that has triggered the process (since the server is answering the whole CR). However, this approach suffers from several drawbacks. Firstly, the users' data is still revealed to a third party (the anonymizer) and thus the problem of preserving the user's privacy has not been solved. That is, we still have no guarantees that the anonymizer cannot be misused if a malicious hacker gains access to it. Secondly, the anonymizer needs to update the current location of all the subscribed users repeatedly, which will require a permanent communication and remote monitoring of the users, which is a clear violation of the users' privacy.

Finally, the robustness of these approaches depends totally on having a relatively big number of neighbors at the time of receiving the requests. Therefore, depending on a middleware is far from being a perfect solution to secure location-dependent queries and hence any secure solution need to communicate directly with the Location Based Server without any intermediate parties.

Kido et al. [KID05] and You et al. [YOU07] have proposed a new technique to hide users' location and trajectory by sending several queries instead of only one. The technique depends on creating several fake queries with fake identities in addition to the real query, thereby, the LBS server will not be able to identify. Apparently, the perfection of this mechanism depends on the number of fake requests generated; the more fake queries generated the more robust and secure the system becomes. The problem with this technique is that as the number of requests sent out by a user grows, the LBS may suspect that it is under an attack and thus the requests may be ignored. Moreover, receiving a big number of requests can slow down the server's response time significantly.

Ghinita et al. [GHI08] have proposed a novel approach based on the Private Information Retrieval (PIR) scheme [GEN05] as well as Grid Cells (GCs). The PIR scheme is used to retrieve data from a database without revealing the content of the queries or the identity of the user. GCs technique is used to request a reduced set of LBS which represents the area of interest to the user. The GCs firstly enquire from the server about the appropriate cells, and then retrieve anonymously suitable objects. Ghinita's technique succeeds in solving some of the issues associated with the abovementioned techniques. However, it relies on unguaranteed expectations like extensive data processing on the user's side. In most cases, the user is submitting the request through a mobile phone that has very limited processing capabilities.

Rebollo-Monedero and Forné [REB10] have proposed a mathematical model to minimize the risk of privacy violations in PIR's queries. They presented a promising system to enhance LBS exchange protocol and make communication more secure, despite using a TTP server as middleware between the users and the LBS server.

## 6.3  Secure location based services

To preserve the privacy of the user while interacting with the LBS server, we present in this section a novel approach based on HES to preserve the privacy of the user while interacting with the LBS server.

As mentioned in previous chapters, HESs allow performing arithmetic operations (additions and multiplications) over encrypted data, meaning that the result of an arithmetic operation would be the same whether applied over plain bits or encrypted bits. Our work uses a symmetric encryption scheme as a basis to request LBS anonymously and guarantee retrieving only suitable data.

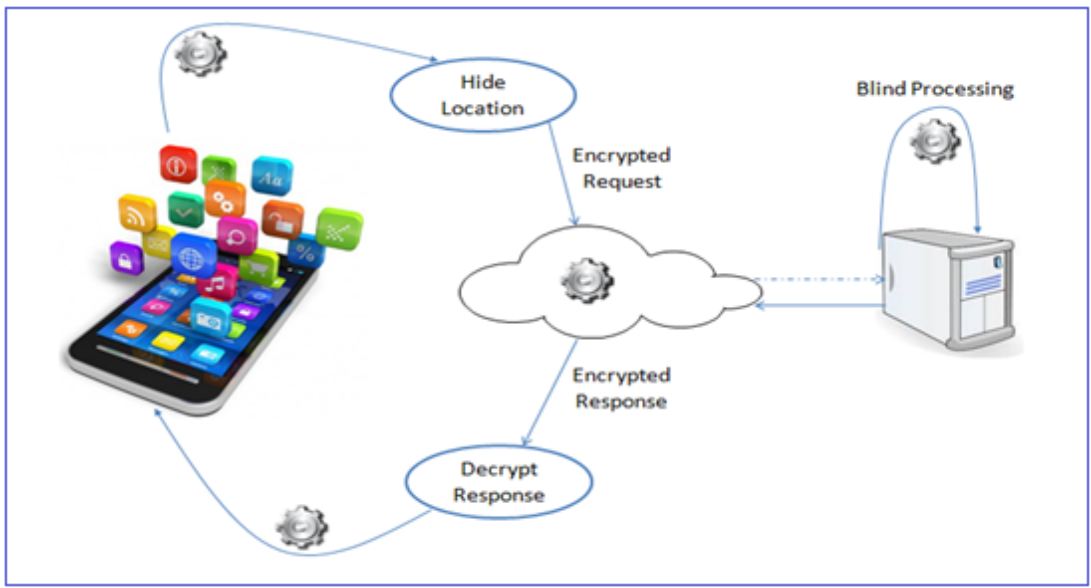

Figure 6-1: Architecture for secure location based services.

Figure 6-1 depicts a high level architecture of the proposed Location Based Service. A user encrypts the request, which consists mainly of the user's geographical position and the category of the service (Bank, University, etc.) he/she is looking for, and then sends the encrypted request

to the LBS server. This latter performs a search on the location database and produces an encrypted result that matches the search criterion. The encrypted records are returned to the user and upon decryption, the requesting party gets the location of the nearest services.

In this approach we use a symmetric encryption scheme to conceive our homomorphic model. Symmetric schemes have the particularity to use the secret key for both encryption and decryption, whereby this characteristic could enhance performance by working on shorter records. This encryption scheme is defined as: $c = m + 2 * r + sk * q$ where $c = \varepsilon_{sk}(m)$ is the cipher text of a bit m encrypted under the secret key $sk$; $r$ and $q$ are two random integers. $2 * r$ is called the noise of the cipher text. By carefully choosing the size of the secret key $sk$ and the random values $r$ and $q$, this encryption scheme is proved to be semantically secure [GEN09a]. The size of each of these elements is based on a security parameter called $\lambda$, whereby $sk$ is an odd $\lambda^5$-bit number, and $r$ and $q$ are $\lambda$-bit and $\lambda^2$-bit numbers, respectively. As a consequence, each bit encrypted using this scheme would be represented in at most $\left(1 + \text{Log}_{10}(Q)\right)$ decimal digits, where $Q$ is $\lambda^7$-bit number. Furthermore, this scheme can support a finite number of arithmetic operations over the same ciphertexts, since it depends mainly on the ratio $sk/r$. Thereby, we are able to decrypt successfully a bit $c$ as long as the noise value is less than $sk$. It is obvious to notice that this value doubles after each addition and squares after each multiplication. Therefore, our proposed process must be carefully executed such that it can terminate tasks before reaching the upper limit of the noise value. It is worth noting that the ability to support a high number of operations depends mainly on the parameter of security $\lambda$. If the latter is relatively large, then the ratio $sk/r$ will be large enough to support a considerable number of computations. However, an encryption scheme that uses a big value of $\lambda$ (to allow a high secure system) produces big encrypted values, whereby the time needed to perform operations will be relatively long. In our system we use the Karatsuba multiplication algorithm [URL01] to manipulate operations over big integer values. This algorithm allows performing more than $10^6$ integer operations in less than one second (tested on a personal computer with 2 GB memory and for the security parameter $\lambda = 5$). This is a practical situation, especially if we consider the size (order of $2^{78125}$) of the encrypted values generated by $\lambda$.

In our system, processing a user's request goes through the following four main steps:

1. Localizing category.

2. Localizing services.

3. Filtering services.

4. Generating results.

We demonstrate how these steps work by the following example. Assuming that the user needs to enquire about the nearby hospitals, he/she sends an encrypted request that represent both his/her current location $(x, y)$ and the enquired category (hospital in this case). Once the server receives the request, it uses the user's position to calculate distances and localize nearby services. Thereafter, it selects only the objects enquired about, and sends them back to the user in encrypted form. We note that the described encryption scheme, allows performing operations between plain and encrypted bits and the resulting record becomes encrypted.

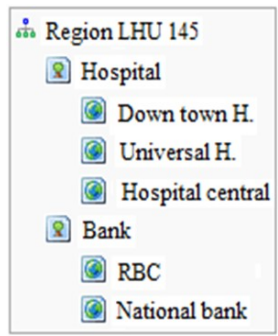

Figure 6-2: LBS's Structure.

In the next sub-sections, we provide complete details on each of these main processing steps.

## 6.3.1 Location categories

The LBS database is structured as a tree, as shown in Figure 6-2. Thus, localizing suitable objects must be preceded by localizing the associated category. This process requires an exhaustive search, since the server doesn't have access to the content of the user's request since it's encrypted. The server needs to compare, bit by bit, the encrypted category, requested by the user, to all available categories in the database. The following formula is used for that comparison:

$$\forall C \in LBS \ \ I_C = \prod_{i=0}^{size-1} [\![ (1 \oplus c_i) \oplus v_i ]\!]$$

Where $size$ is the number of bits used to encode one category, $v_i$ is the $i^{th}$ bit in the enquired category, and $c_i$ is the $i^{th}$ bit in category $c$ that is available in the LBS's database.

This formula focuses on comparing the two categories $c$ and $v$ by verifying whether their sequences of bits are similar or not, knowing that c is encrypted. Towards that end, we compare separately each couple $(c_i, v_i)$ by calculating $1 \oplus c_i \oplus v_i$. The latter results in an encrypted value that either equals to $\varepsilon_{sk}(1)$, if $c_i$ is an encrypted form of $v_i$, or $\varepsilon_{sk}(0)$ otherwise. Then, it is possible to verify whether the compared categories are the same, by checking if all generated values are $\varepsilon_{sk}(1)$. Therefore, we calculate the product $I_C$ of these encrypted bits and if we get $\varepsilon_{sk}(1)$, then the categories are the same. Otherwise, we confirm that at least one pair $(c_i, v_i)$ exists such that $v_i \neq c_i$, meaning that the compared sequences are not the same. The values of $I_C$ are used by the server to filter out the objects that belong to the enquired category. We should finally mention that each entry in $I_C$ will have a noise in the order of $[size \times (\lambda + 1)]$-bit value, since it is the product of size encrypted bits, which are in the order of $(\lambda + 1)$-bit.

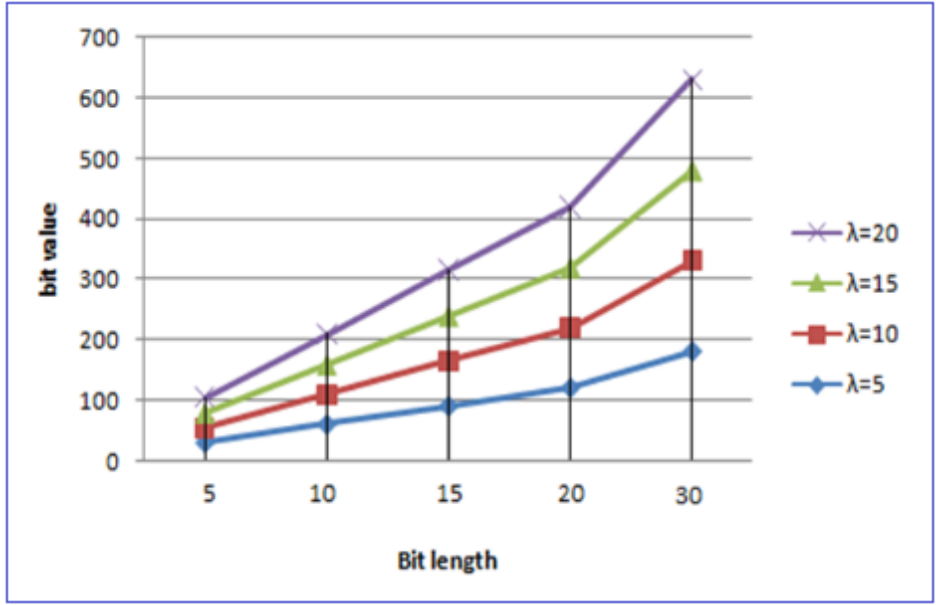

Figure 6-3: Noise value produced while localizing categories.

In Figure 6-3 we show the noise values, produced from localizing categories, with respect to the bit sizes of these categories.

## 6.3.2 Localizing services

In this sub-section we describe the mechanism that locates objects that surround the user's position.

The aim of our approach is to allow users to find the nearest targets while preventing LBS from identifying their positions. Therefore, we use the encrypted position $(\varepsilon_{sk}(X), \varepsilon_{sk}(Y))$ instead of the real one and calculate, based on the Manhattan distance [URL02], the distance separating them from the stored targets.

This distance, depicted in Figure 6-4, is calculated between two points $A = (X_A, Y_A)$ and $B = (X_B, Y_B)$ as follows:

$$d(A, B) = |X_B + (1 \oplus X_A)| + |Y_B + (1 \oplus Y_A)|$$

Where the operations $1 \oplus X_A$ and $1 \oplus Y_A$ represent the two's complement form of $X_A$ and $Y_A$ respectively.

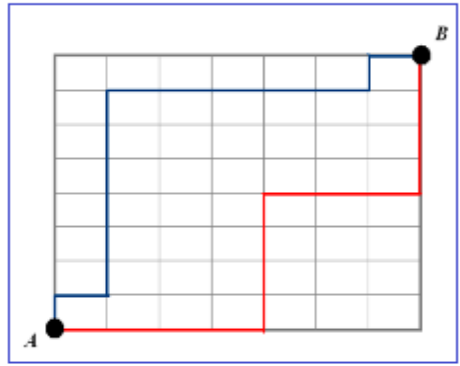

Figure 6-4: Manhattan distance between two points.

The relevant positions are presented as a set of bits, and therefore, binary addition is mandatory to calculate the distance. This arithmetic operation results in noise values ranging from $\lambda$-bit to $(\text{sizeXY} \times \lambda)$-bit values, which are produced when using the same encrypted bits for calculating both the current bit $S_i$ and the carry bit $S_{i+1}$. Here, $\lambda$, sizeXY, and $S_i$ are the original noise, the bit length of the coordinate $(X, Y)$, and the $i^{\text{th}}$ bit in the resulted addition $S$. In other words, this distance produces a noise value confined between $2\lambda$-bit and $(\text{sizeXY}^2 \times \lambda)$-bit, since it is the addition of two sequences with $(\text{sizeXY} \times \lambda)$-bit noise value.

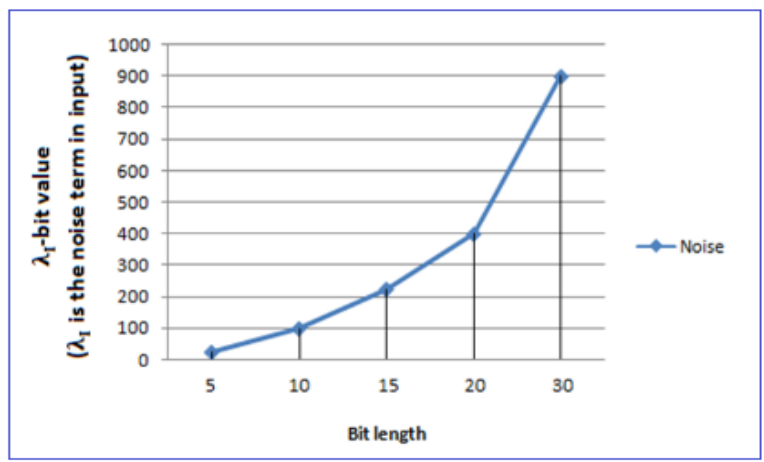

Figure 6-5: Noise value when calculating distances.

In Figure 6-5 we show the noise caused by this step in terms of the coordinate's bit length.

### 6.3.3 Filtering targets

The distance that is separating the targets from the user allows the server to decide which records to suggest for that user. Although the trivial solution is to send back all of the available targets and leave it for the user to filter these targets, this behavior may lead to an excessive processing time and transmission bandwidth. This is because the user must decrypt the results before filtering them which is a time-consuming procedure. Moreover, forcing a user to receive and process many objects irrelevant to his/her search may unnecessarily waste the resources of the user. Therefore, we propose here two novel approaches to mitigate this situation. In one approach, we use a blind sorting process that arranges the targets based on their encrypted distance, and then chooses the closest ones (number of records is known). In the second approach, we blindly localize the points that belong to a coverage area. In what follows we discuss these two approaches, the details about their functionality, and we highlight their benefits and drawbacks.

### 6.3.3.1 Blind sorting

We exploit the PIR methodology and propose a novel circuit that sorts encrypted values. That is, our circuit allows us to arrange the available locations based on their encrypted distance. Our novel circuit uses the principle of blind comparison. This principle compares two sequences of bits, of length n, by performing binary subtraction between them. The $n^{th}$ bit resulting from this subtraction is used to check the nature of the comparison, since this latter is negative whether the bit value is $\varepsilon_{sk}(1)$ and positive otherwise. Blind comparison, however, suffers from a major drawback related to the value of noise that grows rapidly before finishing the process. Therefore, we enhance the model by proposing a novel technique that divides the set of bits into two parts, namely, low and high, and then compares these parts separately, as shown in Figure 6-6.

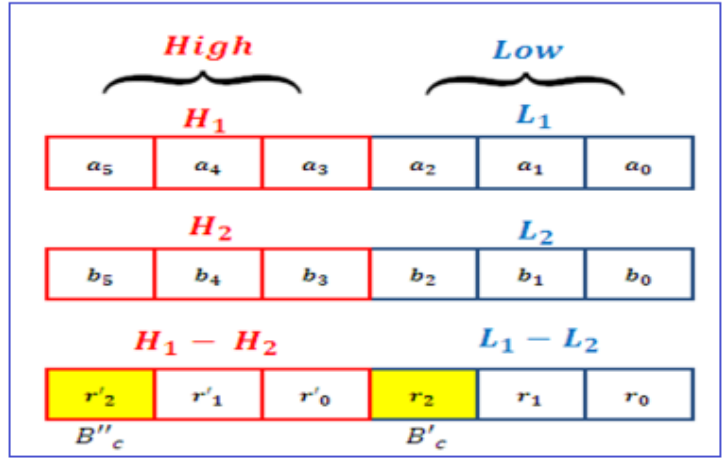

Figure 6-6: Translating the comparison of two set of bits to their low and high parts.

The partial comparisons conducted on the low and high parts of the targeted sequences are used to finalize the comparison between the two sequences using the following equation in order to reset the increased noise value:

$$B_c = B''_c + B'_c \times \overline{B''_c}$$

where, $B_c$, $B'_c$ and $B''_c$ are the resulting encrypted bits of the comparison, the low part, and the high part, respectively.

| $B'_c$ | $B''_c$ | $B_c$ | Order |
|---|---|---|---|
| $\varepsilon_{sk}(1)$ | $\varepsilon_{sk}(1)$ | $\varepsilon_{sk}(1)$ | $\geq$ |
| $\varepsilon_{sk}(1)$ | $\varepsilon_{sk}(0)$ | $\varepsilon_{sk}(1)$ | $\geq$ |
| $\varepsilon_{sk}(0)$ | $\varepsilon_{sk}(1)$ | $\varepsilon_{sk}(1)$ | $\geq$ |
| $\varepsilon_{sk}(0)$ | $\varepsilon_{sk}(0)$ | $\varepsilon_{sk}(0)$ | $<$ |

Table 6.1: Nature of operations from calculated bit value

In Table 6.1 we show the relation between these values, and Figure 6-7 shows how our technique stabilizes the value of the noise, since in both low and high parts we reset the noise value when starting each part.

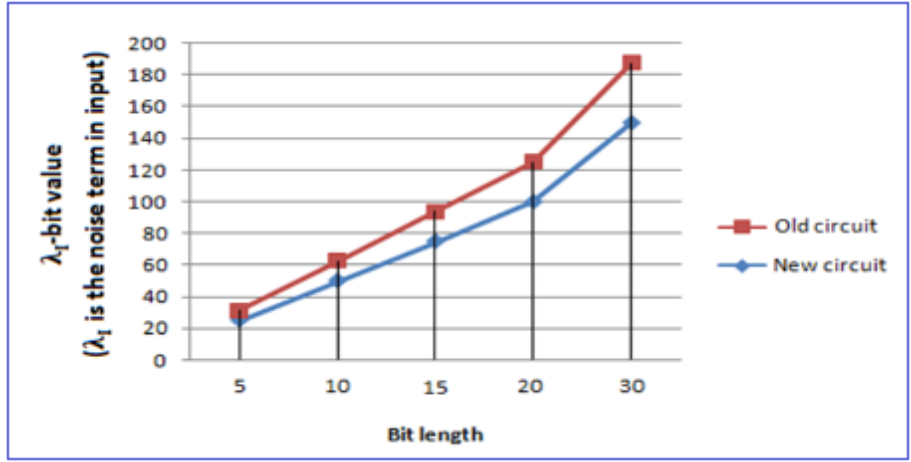

Figure 6-7: Noise value resulting from the new circuit compared to the old one.

Our comparison circuit is then used to arrange the selected objects in a certain order. We compare the two first elements to see if their associated bit $B_c$ is equal to $\varepsilon_{sk}(1)$. If so, we permute these elements, otherwise we keep them in their original order. This process is repeated for each adjacent pair of elements until all of the elements are compared. After that, the full comparison is repeated N times from the beginning, where N is the number of items available in R, the set of targets belonging to a category.

We should mention here that having $B_c$ in an encrypted form may hinder the swap of bits described above. This problem is overcome by using a passkey formula that allows the executor to use the encrypted $B_c$ to apply the comparison. The passkey formula is defined as follows:

$\forall R1 \in R \ and \ R2 \in R :$

$$R1_{new} = B_c \times R1_{old} + \overline{B_c} \times R2_{old}$$

$$R2_{new} = \overline{B_c} \times R1_{old} + B_c \times R2_{old}$$

Where $R1_{old}$, $R2_{old}$ and $R1_{new}$, $R2_{new}$ are the old and new values of the compared entries. Figure 6-8 shows the performance of the latter technique in terms of the noise values it achieves. The technique generates high values of noise because of the fact that the same encrypted values are used $N^2$ times before arranging all of the items.

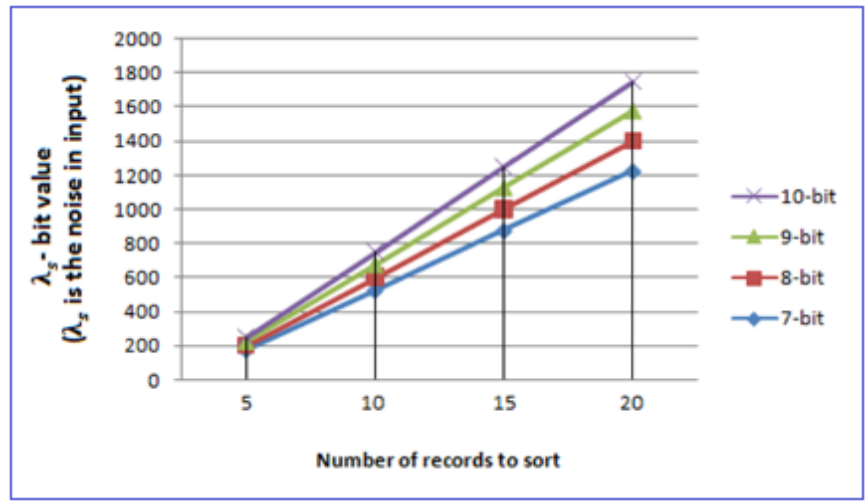

Figure 6-8: Noise value generated from arranging entries.

Therefore, unless the security parameter $\lambda$ is made big enough (i.e., $\lambda > 15$), this circuit does not perform well due to the maximum noise that is reached before finishing the process. For that reason, we propose a lighter technique that focuses on finding the targets that belong to a specific area, instead of arranging all the objects and then choosing the nearest one, in such a way to filter out the suitable entries while keeping reduced noise values.

## 6.3.3.2 Coverage area

Objects can be chosen based on the specific area, surrounding the user's location within a radius P, as shown in Figure 6-9. For that purpose, we compare distances, calculated in the third step to P and then all associated $B_c$ will form a set of encrypted bits called $L_R$. Thus, the location belongs to the specified area only if $L_{R,i}$ is equal to $\varepsilon_{sk}(1)$. Furthermore, each category in the database T will have a corresponding $L_R$ that indicates anonymously whether a target is suitable or not. The fact that the content of $L_R$ is encrypted forces us to find a special process to extract only $L_{R,i}$ that are equal to $\varepsilon_{sk}(1)$. Therefore, we need first to calculate the sum of $L_R$ using the equation:

$$\forall R \in T: S_R = \sum_{i \leq R} L_R$$

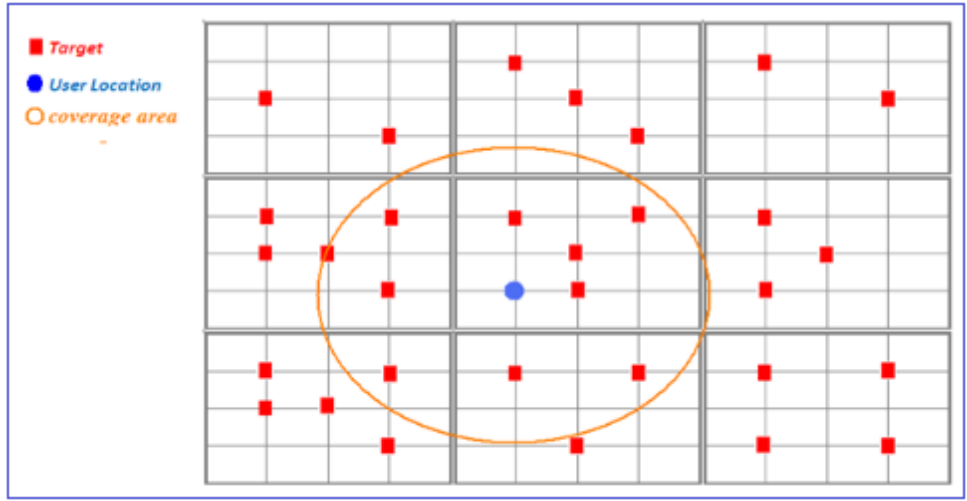

Figure 6-9: Coverage area.

This sum is calculated using elementary symmetric polynomials since this technique keeps the noise value at the order of n, whereby n is the size of $L_R$. It is then possible to localize the $i^{th}$ valid target by calculating the new sequence $L'_R$ as:

$$\forall R \in T \; L'_R = L_R \times \prod_{i=0}^{n}(1 \oplus \eta_i \oplus S_{R,i})$$

Where $\eta_i$ is the binary representation of the index of the element to select. Moreover, $L'_R$ contains only one bit value equal to $\varepsilon_{sk}(1)$ (the index to localize) while all others are equal to $\varepsilon_{sk}(0)$. This sequence leads to constructing a new database $T'$, of rows $R'$, that contains only the targets that belong to the coverage area. The new database $T'$ can be formed as follows:

$$\forall R \in T, \forall R' \in T' : R' = \sum_{i \leq R} (L'_{R,i} \times R_i)$$

Where $R_i$ is the $i^{th}$ target available in the enquired category.

Looking for objects in a specific area produces an acceptable noise value. This is achieved due to the fact that an important number of records can be supported and the circuit can be terminate before reaching the noise's limit. In Figure 6-10 we show the noise produced in this stage.

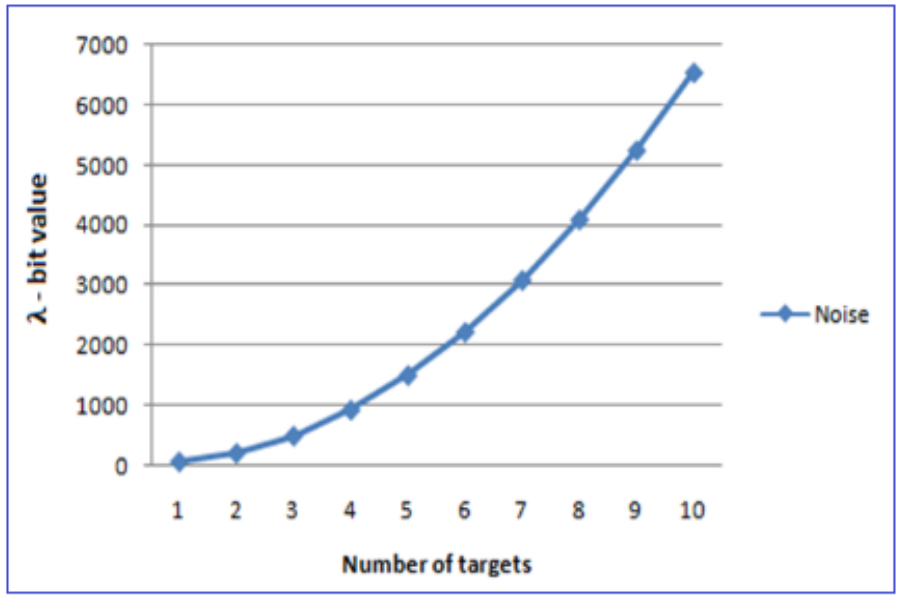

Figure 6-10: Noise value generated while filtering objects in a specific category.

## 6.3.4 Generating results

The last step enables the server to select only the rows $R'$ that belong to the appropriate category. For that purpose we use the sequence of bits $I_C$, that is calculated during the first step, and then multiply it to the locations $R'$ that are available in the fixed perimeter. After that, we calculate the resulting location as follows:

$$\forall R' \in T' : R'' = I_C \times R'$$

## 6.4 Implementation and results

In this section we study the performance of our proposed system. In Figure 6-11 we show the data flow of our proposed system. A user can auto-locate himself/herself (the region where he belongs as well as his position) using smart phone capabilities. Then, the user's software encrypts both his coordinates and the type of services he/she is targeting, and sends them to the server. The server retrieves the requested targets depending on the encrypted information. Thereafter, it sends these encrypted targets to the client to be decrypted and viewed by the user.

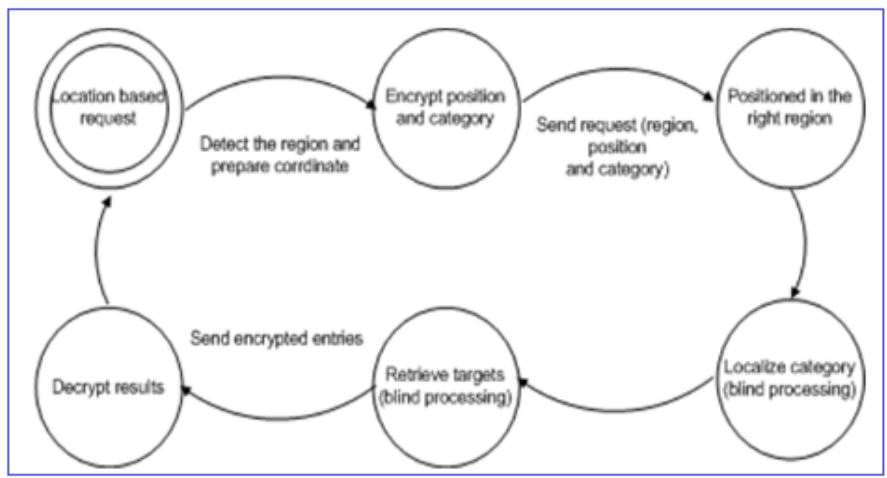

Figure 6-11: Data flow.

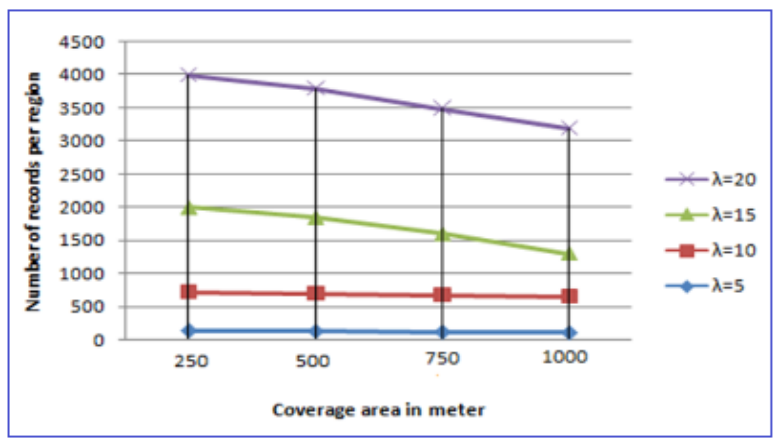

Figure 6-12: Number of records supported for different values of λ.

Our system has a limitation in terms of the number of records it can support. As the number of the stored targets grows, the system needs to conduct a significant number of arithmetic operations. We may even reach the upper limit of the noise value before extracting all the targets, and a successful decryption will not be guaranteed. We can overcome this problem by using large values of security parameter λ.

Our experimental results are shown in Figures 6-12 and 6-13. Our experiments are conducted on a personal computer with 2 GB memory and dual core CPU of 2 GHz. Visual Studio C++ is used as tools for programming.

Figure 6-12 depicts the performance in terms of the number of records per region we can support as a function of λ.

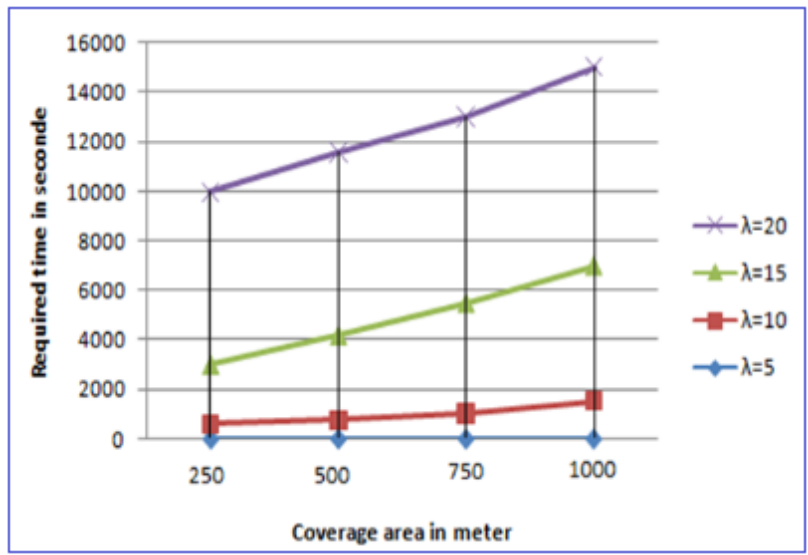

Figure 6-13: Processing time.

Figure 6-13 illustrates the performance in terms of the processing time consumed in each case. This figure shows that whenever the coverage area gets bigger, the number of operations increases, while the number of records supported gets lesser.

## 6.5 Conclusion and Perspectives

The concept of answering location-related information for encrypted positions is promising to improve security needs. Indeed, such a mechanism can strongly attract the attention of researchers as it supports the preservation of the users' privacy.

In this chapter we developed a novel fully secure location-based mechanism based on a HES. We described the circuits that allow a LBS server to process encrypted inputs to retrieve targeted records that match the user's request. We also discussed the limitations and drawbacks of our proposed system and suggested some solutions to make it more practical. The performance of our system was tested through extensive experiments to extract useful results related to the noise generated and the processing time consumed.

As future work, we are planning to improve the performance of the encryption scheme to be able to support a large number of services. This step is mandatory to make it possible for a commercial deployment of our LBS system.

# Chapter 7

# Encrypted trust-based routing protocol

## 7.1  Introduction

Ad-hoc Wireless Network is a set of dynamic nodes connected to each other without any physical liaison or pre-conceived infrastructure. This technology, which is suitable for dynamic topologies, has attracted a valuable attention from scientific community. Whereas, researchers have discussed various aspects to conceive a dynamic strong architecture that matches its counterpart based on some configured parameters. In this area, the routing protocols attract the utmost priority since it focuses on how a packet can be transported over a dynamic network.

The Routing, in an Ad-hoc Wireless environment, relies on the voluntary and collaboration of nodes to forward a packet from source to destination. However, this confidence may sometimes be expensive in a no secure topology, since malicious and selfish nodes may mislead this collaboration and make packets' delivery probabilistic. In a natural way, one prefers to cooperate with only faithful collaborators or at least assigns to them packets with high priority. In this context, an interesting approach has been proposed [PIR04], to avoid selfishness behavior, which is based on a trust mechanism to elect next collaborating nodes. It relies on the evaluation of each node toward its neighbors, where this evaluation affects in-nodes a trust value that is continuously updated according to the degree of cooperation. Trust values are then used to define a trust route. However, many security problems arise in this subject, whereby how neighbors are evaluated is disclosed as well as the exact trust value. Since, no security mechanisms are involved to protect this sensitive information. Therefore, in this approach we focus on developing tracks that allow protecting privacy issues while keeping the same efficiency and robustness. Where, Fully Homomorphic Encryption and Multi-hop Homomorphic concept could be perfectly employed.

FHES are efficient manners to manipulate encrypted data in a blind fashion. Such flexibility encourages publishing encrypted data and delegate safely processing to third parties. In routing purposes this technique is well aligned, since each node can perform its trust circuit and communicate only an encrypted useful evaluation to its neighbors. This way, nodes protect their privacy and intimacy when collaborating in an Ad-hoc network. On the other hand, the FHE concept is bounded by the number of circuits participating on the same computation. Since in the output of a circuit, the resulted ciphertext is no fresher as it was in entry, hence FHE properties might be no more valid. To avoid this limitation, we involve a mechanism called "Adapter" that implements multi-hop concept to preserve FHE properties even with several hops. Such a technique allows us to develop an efficient and correct secure trust based routing protocol for an Ad-hoc wireless network.

The remainder of this chapter is organized as follows. In Section 7.2 we review the related work that aimed at trust routing protocol. Section 7.3 provides an overview of routing techniques as well as their limits. Section 7.4 presents our solution that deals with multi-hop limits and presents our proposition that aims at secure trust based routing protocol, while Section 7.5 details the solution and describes the designed protocol. In section 7.6 we describe briefly the performance of this protocol. Finally, Section 7.7 concludes our work and provides future research directions.

## 7.2 Related work

In this section, we briefly describe other works that aim at define trust mechanisms or aim at secure routing protocol in Ad-hoc wireless environment.

A. A. Pirzada et al. have defined a trust mechanism [PIR04] over a reactive routing protocol. They have proposed a model where each node in an Ad-hoc wireless environment maintains an evaluation procedure to reward or punish nodes in future collaborations. The trust value, which is based on transactions' history and the forwarding quality, is then shared with other nodes to choose a trust path. However, exposing personal experience to a public network threats nodes' privacy and allows malicious module to control the network.

R. K. Nekkanti and C.-W. Lee have proposed in [NEK04] a trust based adaptive routing protocol that denies nodes which are not affected to access route information about routes' path. Furthermore, this contribution has proposed a protocol that uses different kind of encryptions that vary according to the nature of nodes and their trust level. These techniques are first employed to protect route privacy and then to save energy while keeping a smooth functioning.

On one side, S.-H. Chou et al. have adopted in [CHO11] a trust based reputation mechanism for the reactive routing protocol DSR. This contribution has proposed a model that detects malicious and selfish nodes, and then denies them, becoming a part of the protocol. On the other side, K. Sanzgiri et al. have suggested in [SAN02] a secure routing protocol based on authentication mechanisms.

Moreover, C. Zhang et al. have provided in [ZHA10] a formal study of trust based Ad-hoc wireless Network. The proposed work has discussed several subjects regarding this protocol, such as correctness, respecting optimal trajectory to construct paths and distributive of trust branches.

Although the contributions, mentioned above, discussed carefully various aspects that apply trust techniques to routing protocols for a secure matter. None of them has addressed the privacy of this trust model privacy and how this entity should be protected, whereby the way nodes are trusted must be kept confidential as well as trust values.

## 7.3 Trust-Based routing protocol

### 7.3.1 Routing protocol

The wireless Ad-hoc Network is a typical kind of reactive protocols, which has a standalone architecture that does not refer any pre-existed infrastructure to define its topology. As dynamic nodes forming the network, behave in collaborative mode to construct on-demand paths and then route packets node by node until reaching the destination. This kind of network is widely used in a variety of applications and has been introduced in many systems where minimal configuration and quick deployment are required. However, this mechanism suffers from a major drawback concerning quality of routing as well as delivery guarantee. Since the protocol is mainly based on nodes' cooperation and voluntary to announce nodes' availability and broadcast packets hop after hop. Thus, the protocol falls down when a malicious node colludes with other elements to control the network, or when some nodes refuse to collaborate to save energy for instance. Therefore, the selection of nodes that cooperate together should be done carefully, so a protected and safely environment could be provided. This way, several solutions have been proposed to select suitable nodes based on their level of commitment, especially trust based routing technique.

## 7.3.2 Trust-Based routing

The trust concept implied into routing protocols consists of selecting collaborating nodes based on their involvement and the degree of faithfulness. Such factors are extracted using cooperation history and reputation of the nodes, whereby each node believes its neighbors through the level of successful transfer, respect for packages' priority and their availability. A trust value is then assigned to each node by its predecessors, helping to construct a trusted path when a route request is enquired. However, the manner by which nodes are evaluated is not fully protected and a given node can easily extract how it is evaluated and even the exact trust value. Therefore, it is important to hide the criteria used to assess nodes as well as the degree of belief, since these elements are essential to ensure a trust routing and to deny malicious elements to control the network. For this purpose, our contribution aims at protecting nodes' privacy and to allow them sharing their experience safely. This novel technique is based on a FHES and the Multi-hop homomorphic encryption scheme.

# 7.4  Multi-hop homomorphic encryption schemes

## 7.4.1 Fully homomorphic encryption scheme

The Homomorphic encryption is an effective data protection, since a client can only publish encrypted records of his data and then delegate to other parties using and processing these values without decryption. These schemes guarantee a successful decryption after many operations over encrypted circuits. Regardless of problems that the scheme still suffers from (time consuming, limited number of operations, etc), it is considered as a novel way of cryptography that is providing great added values to security issues. Although, some concepts must be developed in order to clarify different sides of this technique and be able to decide what it can and cannot be achieved, such as double encryption, homomorphism using different keys. One of these concepts and which is useful in our contribution is the Multi-hop Encryption Scheme (MHES).

In this chapter, we use the same polynomial public key encryption scheme developed in previous chapters which can be designed as $\varepsilon_{pk}(m) = c = m + 2 * r + pk * q$, where m is a bit value, r a random integer, $q$ the security number and pk the public key. The combination of these elements results c which is an encrypted integer form of m. Homomorphic circuit can then accept encrypted values and successfully compute a set of operations to result another encrypted values which are correctly decrypted only by the party owning the secret key sk, see Figure 7-1. However, in case of multiple circuits acting on the same encrypted data before decryption, see Figure 7-2, requires a specific processing, for that we rely on the MHES to deal with. MHES consists of providing the possibility to process an encrypted entry using different sequenced circuits. However, this issue is not as easy as it appears, since in circuits' output, the ciphertext is no fresher as it was in entry, and by the fact a correct decryption is not guaranteed. Therefore, an alternative way should be proposed to preserve FHE in a Multi-hop environment.

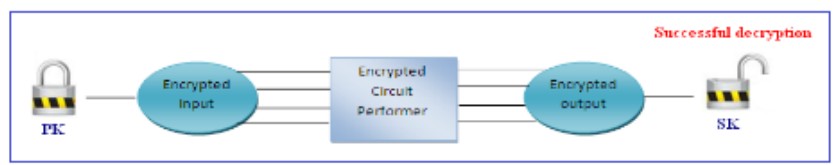

Figure 7-1: Process encrypted values.

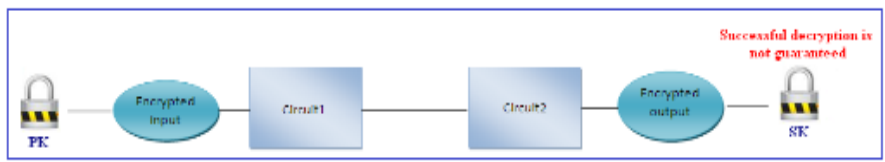

Figure 7-2: Multi-hop encryption.

## 7.4.2 Multi-hop fully homomorphic encryption scheme

Gentry et al. [GEN10b] have discussed this issue in the case of double encryption, and proposed a solution based on the Secure Function Evaluation [SCH08] and the Yao's Garbled circuit [YAO82]. This solution consists of extending the principle of two-party computation to support i-hop HES for a fixed integer i. Recall that two-party computation allows performing a circuit with Heterogeneous inputs. Namely, a server Bob who owns a circuit C can combine his input x (known) with another input y (unknown) to execute its circuit for these values, while ignoring the exact value of y and without revealing the value of its input x. This technique arises as a typical solution of Millionaires' problem that aspires deciding who is the richest without revealing the wealth of both participants. On the other hand, the fact that Gentry relies on such mechanism to fix i-hop theory is discouraging as far as it is promising, since the solution inherits all drawbacks of Yao's garbled circuits, such as pre-calculating the whole circuit as well as transferring from circuit to another the whole truth table, in addition of its current limitations especially time consuming. Despite, the proposed solution is suitable for a topology using many circuits with different public keys.

In this chapter, which consists to secure trust based routing protocol, there is a slight difference compared to i-hop HES discussed by Gentry.

The main difficulty with an environment including several circuits is that a first module $C_1$ acts on the initial fresh ciphertext sent by the client and outputs a processed one. This latter should be given then as input to another module $C_2$ and so on. However, most of times the processed ciphertext does not preserve the same properties than a fresh one, and does not suit exactly what a circuit expects from its predecessor. Therefore, we involve a novel mechanism called

"Adapter" that allows us to transform and customize an output to a set of gates suiting the input of the next circuit if there is one. Namely, when a circuit finishes its processing and before forwarding the output, it checks first how much entries a next node is requiring. Then, it uses the "Adapter" to redraw the output to be exactly what next circuit needs. This way, we form a single chain with several links including all modules that should process client's input, see Figure 7-3.

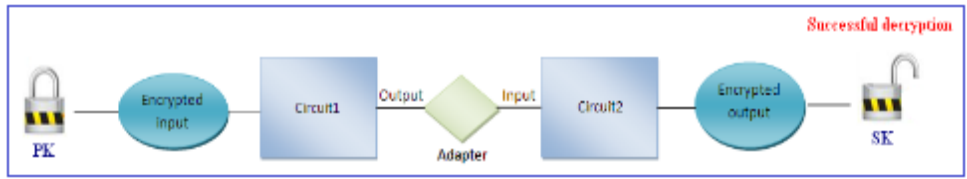

Figure 7-3: Multi-hop architecture using the adapter concept.

### 7.4.3 The Adapter

The adapter concept has resulted from the possibility to transform a set of gates to a new sequence while keeping the same functionality. The new mechanism consists of adapting an output to a next input to form an homogeneous multi-hop HES. However, the "Adapter" concept could be considered as violation of privacy, since an equivalent part of a circuit is shared with its next processor. Thus, such information might disclose important circuits' characteristics. Therefore, we protect privacy issue and nature of circuits by using a third kind of logical gates named "Star". This latter, which has been already proposed the chapter 2, consists of encrypting an ordinary gate to hide whether it is "AND" or "XOR".

Recall, Star gate accepts three encrypted bit values. These values represent the two operands as well as a flag which configures the novel operator to be either a "XOR" or "AND" gate. Thus, a circuit based on star gates refers to a black box that protect circuit privacy, and make from "Adapter" technique a safely one.

To profit from Star gate strengths, we propose to conceive all circuits as many interrelated "Star" gates. Since, this constraint enables these circuits to protect their processing technique and map easily new outputs to next inputs, which realize Adapter concept, without revealing any content. Thus, the "Adapter" concept could not be considered as weakness, since this operation does not disclose any information about the circuit as far as it helps to form one chain.

## 7.5  Encrypted trust-based routing protocol

In this chapter we develop a new mechanism to compute the path between a source node and a destination node, based on the trust metric and the Multi-hop homomorphic mechanism described above, in the goal to provide a secure Ad-hoc Wireless topology for a safely mutual cooperation. We assume that a routing algorithm is used to compute the routes to the destination prior to the establishment of a trusted path to the destination, as well as factors measuring level of faithfulness are already known. In this topology, all nodes maintain their trust database including summary of all transactions with in-neighbors, which is represented as a trust value associated to each collaborating node.

In our protocol the source node begins by generating both public and secret keys, which are used as bases for the Multi-hop HES. This node executes then its trust circuit and calculates trust value that corresponds to each in-neighbor, before selecting among them the most confident to participate in the routing path. Also, it prepares a RR packet carrying the public key, destination node and next hop. Moreover, the current element checks number of inputs required by the next hop, and transforms its output using the "Adapter" to a set of encrypted values. These values represent wires that should be used by the next collaborator, where each three values form one star gate. Thereafter, it maps the generated gates to the RR packet, as well as a field containing encrypted trust value of the most confident next hop. The RR is then forwarded to this latter to continue discovering the trusted path. Each node receiving a RR verifies first whether the destination is among their in-nodes. It proceeds by forwarding the packet without making any changes if this is the case. Otherwise, it begins by dismantling the received packet and extracting the public key and the encrypted accumulated trust value. Thereafter, the node retrieves the set of gates sent by the previous node, and links them to its circuit to continue calculating a trust path. Then, it changes RR content by mapping a trust next hop as well as updating the accumulated trust value for the constructed path as well as the output of "Adapter" mechanism. RR is then transmitted to the concerned node. The protocol follows this way until access the destination node, see Figure 7-4. This latter sends back a RP to the source including the whole path from source to destination, as well as the encrypted trust value associated to the elected path. Finally, the source node is the only actor able to decrypt global trust value, without being aware of the one associated to each arc.

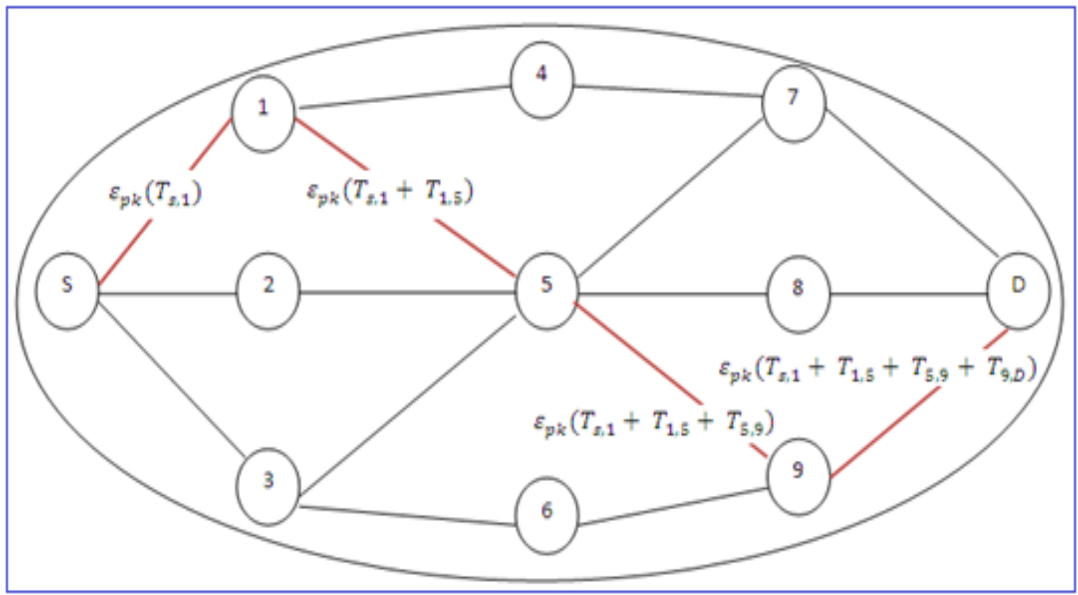

Figure 7-4: Accumulate encrypted trust value T from source to destination.

## 7.6  Evaluation and results

In this section we discuss the performance of the proposed protocol and its impact to an ordinary trust based mechanism. The performance of our system is mainly based on the parameter of security $\lambda$, the bigger this parameter is the more secure the system becomes. In return, the system will deal with huge operations when using big values of $\lambda$.

We study the case of an Ad-hoc network with 20 different nodes where we should select a trusted path and accumulate blindly the trust value associated. Each node in this topology first needs to evaluate its in-neighbors, and then updates the current trust value before forwarding the packet. This trust value, varying from 1 to 10 to represent the lower and the higher level of trust respectively, can be represented in 4 bits. While its encrypted form is at order of $4 \times \lambda^5$ bits. On the other hand, each node participating in the protocol should perform 20 additions and 8 multiplications (homomorphic operations over encrypted values) to update the previous trust value. Meaning that to calculate a path in a network of 20 nodes we need to compute at most 400 additions and 160 multiplications.

We note that for faster computations, we rely on Karatsuba algorithm [URL01] which is suitable for operations over big integers. This algorithm allows performing more than $10^9$ integer operations in less than one second (tested on a PC with a 6 GB of RAM memory at a security parameter $\lambda = 3$).

We show in Table 7.1, the needed time to compute an encrypted trust path according to different value of $\lambda$ in a network of 20 nodes.

| Security Parameter λ | Needed time in seconds |
|:---:|:---:|
| 3 | 0.3 |
| 5 | 1 |
| 8 | 3 |
| 10 | 8 |

Table 7.1: Needed time to compute the encrypted trust path.

## 7.7 Conclusion and perspectives

In this chapter, we aimed at developing a secure trust based routing protocol using Multi-hop homomorphic encryption concept. This contribution consists of two main parts; the first one is proposing a model that realizes Multi-hop encryption in an environment using one encryption layer. This solution is then used to conceive a secure routing protocol, which bases on the personal experience and the valuation of each node regarding the faithfulness of their neighbors, to define a trust path. Moreover, protecting the privacy and the evaluation level of these nodes is guaranteed, since the solution is based on a FHES. The proposed technique allows nodes' collaborating and sharing experiences in a public topology without disclosing sensitive information.

As future work, we plan to describe a fully scheme that hides the source and the destination while computing the trusted path.

# Chapter 8

# A fully private Video on-Demand service

## 8.1 Introduction

The popularity of VOD services mandates the need to preserve both the confidentiality of information and the privacy of the clients. From one side, service providers seek to protect their video store by allowing access for authorized clients only. On the other side, authorized clients require that service providers gain no knowledge of the materials they purchase or watch (to prevent any kind of tracking or monitoring). This means that we need to implement secure protocols that prevent service providers and clients from extracting or accessing important information about each other. Such a protocol can be realized by the use of HESs. In these schemes it is possible to perform arithmetic operations directly on encrypted operands. That means, this kind of scheme enable blind processing, and finds applications in many areas where confidentiality and privacy are of major concerns. Securing video streams transported over the Internet can be an important area to benefit from the strengths of HESs.

In this chapter we propose an efficient protocol that relies on HESs to provide a secure and private video on-demand service. The proposed protocol allows users to purchase videos from networks with low security measures (privacy is not protected). This idea is to encrypt users' queries and blindly (that is, without being aware of their contents) process/execute them, see Figure 8-1. We also provide an analytic study to describe the protocol's data transfer and recognize the bandwidth needed to support the whole architecture.

The rest of this chapter is organized as follows. In Section 8.2, we review the work related to the area of secure video streaming. Section 8.3 describes and details the concept of HESs and presents our new protocol. Section 8.4 provides an analytic study and evaluates the performance of the new protocol. Finally, Section 8.5 concludes our work and provides future research directions.

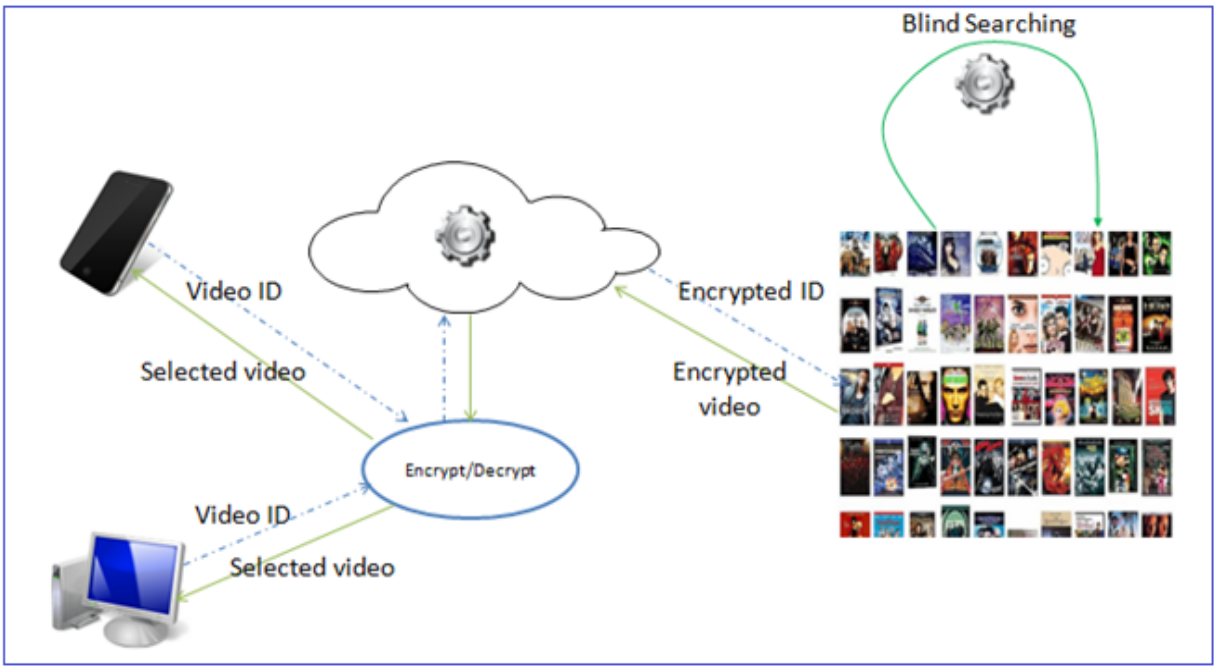

Figure 8-1: A fully secure VOD architecture.

## 8.2  Related Work

In this section, we review several research contributions that aim at securing video streaming over public networks.

Yin et al. [YIN05] develop a mechanism that combines Content Delivery Network and P2P to achieve secure and scalable media exchange architecture. This mechanism uses a selective platform that elects two members, namely, the cluster leader and the backup leader. As the streaming environment is dynamic, these two members accept or reject new incoming elements forming the group and adapt the network according to changes. Also, the mechanism protects media contents by relying on two encryption keys: the session key (SK) and the cluster key (CK). The SK encrypts the flow while the CK encrypts the SK itself before forwarding it over a secure channel. Both of these keys are then used by authorized members to decrypt streaming contents.

Bachtiar et al. [BAC95] proposed an encryption mechanism, called the bit-serial cryptosystem, which can be used to support security in video on-demand systems. Using a pseudo-random Shift Register, the mechanism can be used to efficiently encrypt and decrypt stream signals. Venkatesh et al.[VEN09] use cryptographic approaches to protect privacy and integrity in a video conference system. The technique first relies on a pitch shifting mechanism in which the audio stream is simply modified, and then it uses a video obfuscation to ensure efficient privacy.

Raju et al.[RAJ08] choose RC5 to protect video privacy. While RC5 is a block-cipher encryption technique, the authors have proposed dividing each frame into several blocks, and then apply the encryption algorithm with a key size of 128 bits to cipher video stream as text data.

Chen and Lee [CHE05] provide a solution that encrypts only important set of bits and not the whole sequences. Such a solution can reduce computing time without compromising encryption efficiency.

Lianet. et al. [LIA06] use selective techniques that encrypt the videos partially (that is, encrypting the entire compressed video stream is avoided). While the contribution in [LIA06] relies on AVC codec, the data stream is encrypted segment by segment to ensure an efficient mechanism. Liu and Koenig [LIU05] use the puzzle game principle to introduce an interesting technique for video encryption. The technique suits high resolution streams.

Sun et al. [SUN06] propose a content-based authentication solution for video trans-coding with which several streaming environments are supported. This solution recognizes the fact that streaming data should be digitally converted to support diverse requirements, such as bandwidth and terminal capacity.

Du et al. [DU08] propose a solution that is based on Virtual Private Network technology to support a secure conference system.

Bertino et al. [BER00], Chen et al. [CHE08], and Moon et al. [MOO09], introduce new kind of architectures that control video stores and protect video contents against security violations from different attacks.

It is worth noting that none of the above mentioned techniques provide a solution to protect both confidentiality and privacy at the same time. This observation motivates the research efforts towards devising a system that can address these security aspects together. This system can achieve a fully secure and private video streaming. With this system, a client can request videos privately without having the service provider reveal the nature of that request. On the other hand, the client can have access only to authorized videos such that the service provider protects its store.We aim at devising security solutions that target this kind of systems. Our work is based on the fully HES proposed in [GEN09a].

## 8.3  Secure and private stream protocol

The Internet network was mainly built to facilitate the exchange of information among different entities. Security measure did not receive a strong attention in that design. This means that the regular usage of Internet like browsing, watching videos, and research are not sufficiently secured to protect the privacy of the users. The extensive researches in the areas of data warehouse and data mining have shown that these regular users' activities are very dangerous. This is because these activities can be utilized by intruders to track users' habits and even retrieve their personal information.

In this section we present a secure architecture that can improve the security of public stream protocols. Our main focus is on the video on-demand platform.

In what follows we will reuse the FHE polynomial scheme : $\varepsilon_{pk}(m) = c = m + 2 * r + pk * q$ while we choose pk and q as $\lambda^2$ bit value.

To preserve clients' privacy when interacting with video stores over public networks (like the Internet), we propose the following encryption technique. Firstly, a client checks the video store, which keeps its videos arranged by identifiers (IDs), and selects a video. Then, the client fixes the parameter of security $\lambda$, and invokes the key generator to produce a public key, for encryption, and a secret key, for decryption. After that, the client encrypts the selected video's identifier ID, which is encoded under 10 bits, bit by bit as follows:

$$ID_{pk,i} = ID_i + 2 * r + pk * q$$

Where $ID_i$ is the $i^{th}$ bit in the binary form of value $ID$, and $ID_{pk,i}$ is its encryption form under the generated public key $pk$. These encrypted values are then sent as an encrypted request to the provider that is supposed to stream the suitable video. When the provider receives the encrypted request, it proceeds to find the desired record. This process requires an exhaustive search as the provider does not know the content of the request. Therefore, the provider is forced to compare, bit by bit, the encrypted video identifier, requested by the client, to all available videos in the store. The following formula is suitable for that comparison:

$$\forall V \in STORE \ \ I_v = \prod_{i=0}^{size-1} \llbracket \big(1 \oplus ID_{pk,i}\rrbracket \oplus v_i \big)$$

Where size is the number of bits used to encode the video identifiers, $ID_{pk,i}$ is the $i^{th}$ bit in the enquired category, and $v_i$ is the $i^{th}$ bit in the video V which is available in the store.

This formula focuses on comparing two identifiers blindly by checking whether their sets of bits are the same or not. This comparison results in $\varepsilon_{sk}(1)$ if the identifiers are similar or $\varepsilon_{sk}(0)$ otherwise. In the exhaustive search through the videos in the store we compare each video to the request by calculating $1 \oplus ID_{pk,i} \oplus v_i$. This search generates a sequence of encrypted bits $I_v$, where the $i^{th}$ item contains the comparison result of the $i^{th}$ video to the query. It is worth noting that the provider's store maintains the data in an unencrypted format. That is, the afore mentioned comparison is performed between an encrypted set of bits and plain ones. However, irrespective of this fact, the scheme is still convenient since it supports computing over heterogeneous values, encrypted and plain.

To retrieve the targeted video and ignore other ones, we apply this formula to the store:

$$\forall\, V \in STORE : R = \sum_{i \leq V} (I_{v,i} \times V)$$

Where R is the encrypted stream flow resulted and requested by the client. This formula uses the comparison results to filter out the targeted video by multiplying each element in $I_v$ to its associated video. This way, the confidentiality and integrity of the provider are also guaranteed since we calculate as many stream flows as there are videos available in the store. Thereby, only one flow matches the client's request, while the other flows are just set of $\varepsilon_{sk}(0)$. Thus, a simple bit-by-bit addition between all of the flows satisfies the query. The latter is sent back to the client in an encrypted form. The client can decrypt the received data using his/her secret key and then extract the content.

From the above description we can see that both sides of the communication (the client and the service provider) can preserve their privacy. From one side, the client hides his intentions and can still benefit from the provider's services. On the other side, the provider grants the client an access to only authorized products.

## 8.4 Evaluation and results

In this section we evaluate the performance of the proposed system. The performance is mainly dependent on the parameter of security $\lambda$; the bigger this parameter is the more secure the system becomes. However, the length of this parameter has a major impact on the system performance. That is, as the length increases, the operations of the system start to deal with big values (at the order of $2^{\lambda^4}$ integer value) and this is highly time-consuming. Therefore, this security parameter should be carefully configured to balance between the achieved security level and the delay in the system.

Furthermore, we note that for an optimal performance, we rely on Karatsuba algorithm [URL01] to have faster computations over big integers. This algorithm allows performing more than $10^9$ integer operations in less than one second (tested on a PC with a 6 GB of RAM memory at a security parameter $\lambda = 3$).

We study the transfer of a video of size 100 Mo (10 minutes of record). Also, we suppose that the bandwidth for the normal video stream is 2.5 Mbit/Second. The video ID, which has a plain value at the order of $2^{10}$ bit, is firstly encrypted by the client. This operation results in an encrypted ID in which each bit is transformed into a $2^{\lambda^4}$ integer value at most. Thus, for an ID of 10 bits, we have $10 \times \lambda^4$ serial bits to transfer to the provider. The provider uses the transferred data to select and implicitly encrypt one plain video content. This implicit encryption is needed for the comparisons performed between the encrypted ID and the plain data stored at the provider's store. After finding the targeted video, the size of the encrypted content of this video is almost equal to VideoSize $\times$ EncryptionBitSize (that is, $10^8 \times \lambda^4$ bits).

On the other hand, this comparison is performed after $10^{14}$ additions and $10^{11}$ multiplications, producing a processing time varying from 2 seconds to 12 seconds for $\lambda = 3$ and $\lambda = 6$, respectively.

In Table 8.1, we see the encrypted size as well as the required bandwidth to correctly stream the encrypted flow for different values of $\lambda$. These results is performed through the following formula:

$$\text{EncryptedStream} = \text{OrginalStream} \times \lambda^4$$

$$\text{RequiredBandwidth} = \text{OrginalBandwidth} \times \lambda^4$$

| Security Level | Original Video SIZE (In Mega bytes) | Video SIZE using homomorphic scheme (In Mega bytes) | Bandwidth Required for streaming encrypted video |
|---|---|---|---|
| 3 | 100 | 8100 | 202 Mega/Second |
| 4 | 100 | 25600 | 640 Mega/Second |
| 5 | 100 | 62500 | 1562 Mega/Second |
| 6 | 100 | 129600 | 3240 Mega/Second |

Table 8.1: Encrypted stream size and bandwidth required for a video of 100 Mega byte.

In Figure 8.2 we see the bandwidth value for different values of $\lambda$.

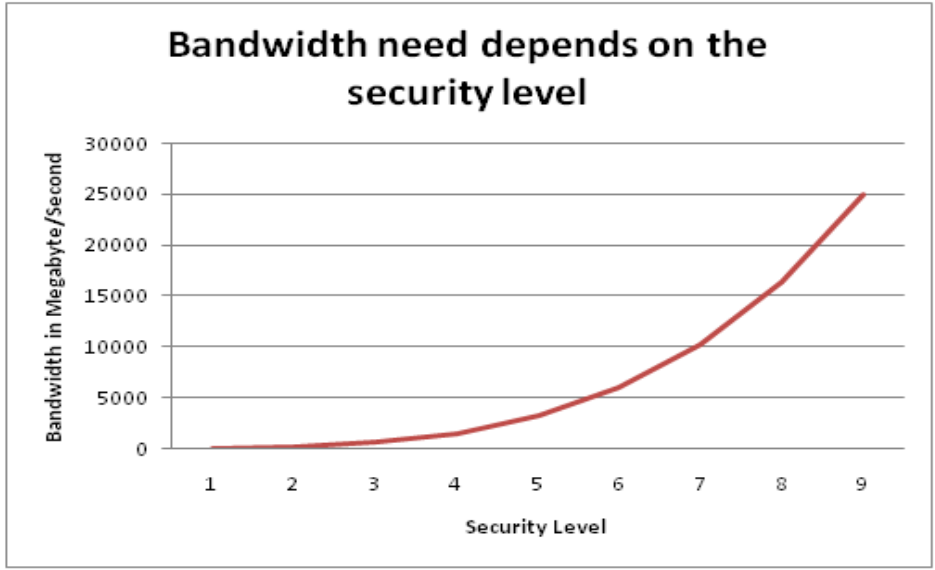

Figure 8-2: The bandwidth required for different security level.

Furthermore, as the stream content increases after encryption, the cache storage increases as well. Then, we calculate the amount of cache as:

$$\text{Storage (MB)} = \text{length (s)} \times \text{bandwidth (bit/s)} / 8 \times 1024 \times 1024$$

In Figure 8-3, we show the amount of storage needed per λ.

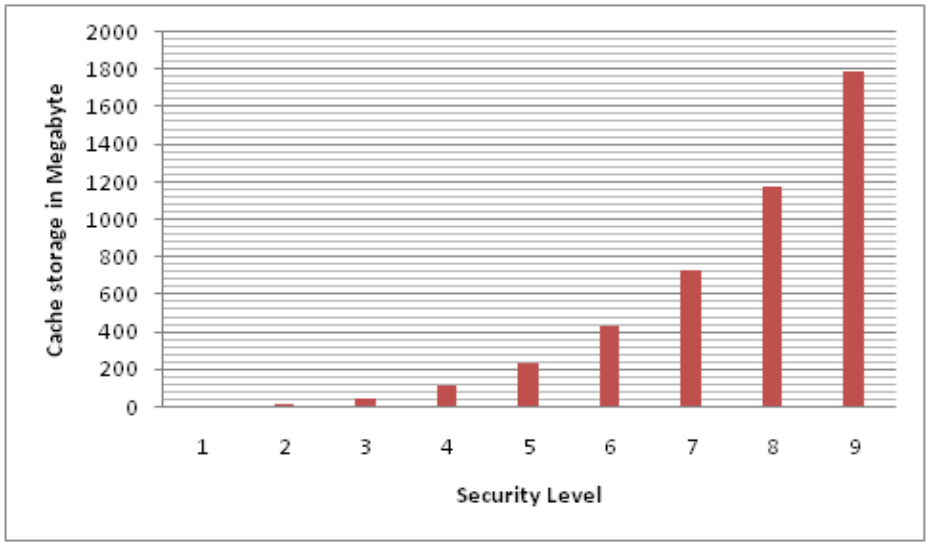

Figure 8-3: The bandwidth required for different security level.

Indeed, the encrypted stream is $\lambda^4$ time bigger than the original one. However, the encountered delays in our proposed system are not of a major concern, given the outstanding advancements in networking technologies, especially the emergence of the Long Term Evolution (LTE) that supports up to 100 Mega bytes per second.

## 8.5 Conclusion and perspectives

The secure stream transfer over a public network (like the Internet) is a promising concept that has major important applications. With this concept, clients' privacy is preserved such that the tracking of their Internet usage can be effectively prevented.

In this chapter, we have provided a secure protocol, based on a FHES that balances between clients' interests and those of the service provider to form a private and secure video on-demand service.

We have described how clients prepare queries, and how the provider blindly processes these queries without accessing their contents. We have shown how our protocol highly depends on the security parameter $\lambda$ and how the configuration of this parameter affects the system's performance in terms of delay. However, the increased delay may not be a major problem given the high speeds supported by the latest technologies (like LTE).

As a future work, we plan to improve the fully homomorphic scheme to support the processing of higher number of nodes.

# Chapter 9

# Conclusions and Future Research

## 9.1  Concluding Remarks

In this thesis, we have tackled the problems of secure applications, services and routing protocols. We have relied on the HES to design efficient circuits that are suitable to fully protect user privacy, and capable of securing the protocols of communication. These circuits have been shown to be customizable and implementable in various applications and domains. In this thesis we had the following achievements:

1.  We have re-designed basic database queries using FHES.

2.  We have presented an encrypted processor that allows executing programs without revealing their details and internals.

3.  We have built a secure architecture for Location Based Services to demonstrate the effectiveness of our novel approach.

4.  We have developed a secure mechanism for information exchange in ad-hoc wireless networks.

5.  We have developed a fully secure mechanism that establishes a secure protocol between video requesters and service providers.

However, these schemes are developed for the "honest-but-curious" adversary model. This model assumes that the interveners work in a collaborative mode, even if the harm they cause is as simple as data alterations. Furthermore, the performance of these schemes remains under the scope of the research interest as they should be enhanced to support more computations. Finally, the adopted FHE actually suffers from the finite number of operations, which requires a re-encryption procedure to support an arbitrary number of operations. The latter procedure contributes to an extra deterioration of the performance.

## 9.2  Future Research

1. The proposed circuits can be extended to face various attacks, even if the server does not collaborate faithfully. Also, these models should be able to check whether the performer executes the requests as they are or not.

2. The used scheme can be improved to enable performing the circuits within a reasonable time. The improvements may consider enhancing the keys generator by providing a fast and efficient system, reducing the length of the ciphertext without undermining the robustness, and bounding the noise value after performing the additions and multiplications.

3. While re-encryption is strongly required to support an increased number of operations, an efficient protocol to refresh the ciphertext will have a great added value.

4. The adopted techniques in this thesis rely on the mono-setting architecture, which means that we consider only a single user and a single server. However, the multi-setting topology that supports several users, that have access to one shared area, is closer to the reality.